\documentclass[usenatbib]{mn2e}
\usepackage{graphicx}
\usepackage{mathptmx}
\usepackage{longtable}
\title[A blind detection of complex SZ structure]{A blind detection of a large, 
      complex, Sunyaev--Zel'dovich structure\thanks{We request that any 
      reference to this paper cites `AMI Consortium:Shimwell et~al.\ 2010'}}

\label{firstpage}

\author[Shimwell et~al.]{AMI Consortium:
 T. W. Shimwell$^1$\thanks{E-mail: tws29@mrao.cam.ac.uk},
 R. W. Barker$^1$, 
 P. Biddulph,
 D. Bly,
 R. C. Boysen,
 \newauthor
 A. R. Brown,
 M. L. Brown$^{1,5}$,
 C. Clementson,
 M. Crofts$^1$,
 T. L. Culverhouse$^2$,
 J. Czeres,
 \newauthor
 R. J. Dace,
 M. L. Davies$^1$,
 R. D'Alessandro$^1$,
 P. Doherty$^1$,
 K. Duggan,
 J. A. Ely$^1$,
 M. Felvus,
 \newauthor
 F. Feroz$^1$,
 W. Flynn,
 T. M. O. Franzen$^1$,
 J. Geisb\"usch$^3$,
 R. G{\'e}nova-Santos$^4$,
 \newauthor
 K. J. B. Grainge$^{1,5}$,
 W. F. Grainger$^6$,
 D. Hammett$^1$,
 M. P. Hobson$^1$,
 C. M. Holler$^8$,
 \newauthor
 N. Hurley-Walker$^1$,
 R. Jilley,
 T. Kaneko,
 R. Kneissl$^{8}$,
 K. Lancaster$^{10}$,
 A. N. Lasenby$^{1,5}$,
\newauthor
 P. J. Marshall$^{9,11}$,
 F. Newton,
 O. Norris,
 I. Northrop$^1$,
 D. M. Odell$^1$,
 M. Olamaie$^1$,
 \newauthor
 Y. C. Perrott$^1$
 J. C. Pober$^{12}$,
 G. G. Pooley$^1$,
 M. W. Pospieszalski$^{13}$,
 V. Quy,
\newauthor
 C. Rodr{\'i}guez-Gonz{\'a}lvez$^1$,
 R. D. E. Saunders$^{1,5}$,
 A. M. M. Scaife$^{1,14}$,
 M. P. Schammel$^1$,
\newauthor
 J. Schofield,
 P. F. Scott$^1$,
 C. Shaw$^1$,
 H. Smith$^{9}$,
 D. J. Titterington$^1$,
 M. Veli{\'c},
 \newauthor
 E. M. Waldram$^1$,
 S. West,
 B. A. Wood,
 G. Yassin$^{9}$ and
 J. T. L. Zwart$^{15}$\\
}

\date{Accepted ---; received ---; in original form \today}

\pagerange{\pageref{firstpage}--\pageref{lastpage}}

\pubyear{2010}
\begin{document}

\maketitle

\begin{abstract}
We present an interesting Sunyaev--Zel'dovich (SZ) detection in the
first of the Arcminute Microkelvin Imager (AMI) `blind',
degree-square fields to have been observed down to our target
sensitivity of $100 \mu \rm{Jy} /beam$. In follow-up deep pointed
observations the SZ effect is detected with a maximum peak decrement
greater than 8 $\times$ the thermal noise. No corresponding emission
is visible in the ROSAT all-sky X-ray survey and no cluster is evident
in the Palomar all-sky optical survey. Compared with existing SZ
images of distant clusters, the extent is large ($\approx 10'$) and
complex; our analysis favours a model containing two clusters rather
than a single cluster.  Our Bayesian analysis is currently limited to
modelling each cluster with an ellipsoidal or spherical beta-model,
which do not do justice to this decrement. Fitting an ellipsoid to
the deeper candidate we find the following. 
(a) Assuming that the \cite{Evrard_02} approximation to \cite{Press_Sch} 
correctly
gives the number density of clusters as a function of mass and
redshift, then, in the search area, the formal Bayesian probability ratio of the AMI
detection of this cluster is 7.9 $\times$ $10^{4}$:1; alternatively
assuming \cite{Jenkins} as the true prior, the formal Bayesian probability ratio of
detection is 2.1 $\times$ $10^{5}$:1. (b) The cluster mass is
${M_{T,200}}=5.5^{+1.2}_{-1.3}$$\times$ $10^{14}
{h_{70}^{-1}M_{\odot}}$.  (c) 
Abandoning a physical model with number density prior and instead simply
modelling the SZ decrement
using a phenomenological $\beta$-model of temperature decrement as a function of angular distance, we find a central SZ temperature decrement
of $-295^{+36}_{-15}$~$\mu$K -- this allows for CMB primary
anisotropies, receiver noise and radio sources. We are unsure if the cluster
system we observe is a merging system or two separate clusters.
\end{abstract}


\clearpage

\bgroup\small\it\ignorespaces\raggedright
$^1$ Astrophysics Group, Cavendish Laboratory, J J Thomson Avenue, Cambridge CB3 0HE\\
$^2$ Owens Valley Radio Observatory, California Institute of Technology, Big Pine CA 93513 \\
$^3$ National Research Council Canada, Herzberg Institute of Astrophysics, Dominion Radio Astrophysical Observatory, P.O. Box 248, Penticton, BC, V2A 6J9, Canada \\
$^4$ Departamento de Astrof\'{\i}sica, Universidad de La Laguna, E-38205 La Laguna, Tenerife, Spain \\
$^5$ Kavli Institute for Cosmology Cambridge, Madingley Road, Cambridge, CB3 0HA\\
$^6$ School of Physics and Astronomy, Cardiff University, Cardiff, CF24 3AA \\
$^7$ Hochschule Esslingen, Kanalstrae 33, Esslingen 73728, Germany \\
$^8$ Joint ALMA Office, Av El Golf, 40, Piso 18, Santiago, Chile \\
$^9$ University of Oxford, Denys Wilkinson Bldg, Keble Road, Oxford OX1 3RH \\
$^{10}$ H. H. Wills Physics Laboratory, University of Bristol, Tyndall Ave, Bristol BS8 1TL \\
$^{11}$ Kavli Institute for Particle Astrophysics and Cosmology, Physics Department, Stanford University, Stanford, CA, 94305, USA \\
$^{12}$ Department of Astronomy, University of California at Berkeley, Berkeley, CA 94720 \\
$^{13}$NRAO Technology Center, 1180 Boxwood Estate Road, Charlottesville, VA 22903, US \\
$^{14}$ Dublin Institute for Advanced Studies, 31 Fitzwilliam Place, Dublin 2, Ireland \\
$^{15}$ Columbia Astrophysics Laboratory, Columbia University, 550 West 120th Street, New York 10027, USA\\
\egroup

\section{Introduction}

The Sunyaev--Zel'dovich (SZ) effect is the inverse-Compton scattering of cosmic
microwave background (CMB) photons from the hot plasma within clusters of
galaxies (\citealt{SZE}, see e.g. \citealt{BIRK_SZ_REVIEW} and
\citealt{CARL_SZ_REVIEW} for reviews). The surface brightness of an SZ
signal does not depend on the redshift ${z}$ of the cluster and the integrated signal is only weakly dependent on $z$ via the angular diameter distance. Hence an SZ-effect
flux-density-limited survey can provide a complete catalogue of galaxy clusters
above a limiting mass (see e.g. \citealt{BART_SILK},
\citealt{AMI_EXPECTED_RESULTS}, \citealt{ACT_KOSO} and \citealt{SPT_INTRO}).

Detecting and imaging the SZ effect has gradually become routine since it was
first securely detected by \cite{Birkinshaw_1981} and first imaged by
\cite{Jones_1993}. Until recently, SZ observations have been directed almost
entirely towards clusters selected optically or in X-ray, for example with AMI
(\citealt{7CLUSTERS}), AMiBA (\citealt{AMiBA_WU}), APEX (\citealt{APEX_HALV}),
CBI (\citealt{CBI_UDOM}), CBI-2 (\citealt{CBI_PEAR}), OCRA
(\citealt{OCRA_LANC}), OVRO/BIMA (\citealt{CARL_JOY}), RT
(\citealt{RT_GRAINGE}), SuZIE (\citealt{SuZIE_HOL}), SZA (\citealt{Muchovej_2011}) and the VSA
(\citealt{VSA_LANC}). Now, however, SZ blind surveying is underway, with ACT
and SPT having produced initial results (\citealt{ACT_CLUSTERS},
\citealt{ACT_CLUSTERS2}, \citealt{SPT_2009}, \citealt{SPT_CLUSTERS2} and
\citealt{SPT_CLUSTERS3}). The Arcminute Microkelvin Imager (AMI) is conducting
a blind cluster survey at 16\,GHz in twelve regions, each typically one
deg$^2$, which contain no previously recorded clusters. The AMI cluster survey
focuses on depth, aiming to detect weak SZ-effect signals from clusters of
galaxies with a mass above $M_{T,200}$ $=$ 2 $\times$ $10^{14}$$M_{\odot}$, where
$M_{T,200}$ corresponds to the total cluster mass within a spherical volume such that
the mean interior density is 200 times the mean density of the Universe at the
current epoch.

The outline of this paper is as follows. In Section \ref{sec:OBS_INFO}, we give a brief description of the instrument, observations, data reduction and map making techniques. Identifying cluster candidates is described in Section \ref{sec:METHOD} -- we stress that some readers will wish to jump to the start of Section \ref{sec:METHOD} which is an important overview of the three analysis methods  and of their assumptions. We discuss how we apply a Bayesian analysis to the AMI data in Section \ref{sec:BAYES} and present the results in Section \ref{sec:RESULTS}.

We assume a concordance $\rm{\Lambda}$CDM cosmology, with
$\rm{\Omega_{m}}$ = 0.3, $\rm{\Omega_\Lambda}$ = 0.7 and H$_{0}$ = 70
km\,s$^{-1}$Mpc$^{-1}$. The dimensionless Hubble parameter $h_{70}$ is defined
as $h_{70}$ = H$_{0}$/(70 km\,s$^{-1}$Mpc$^{-1}$). All coordinates are given at
equinox J2000.

\section{Instrument, observations, data reduction and mapping}\label{sec:OBS_INFO}

\subsection{The Arcminute Microkelvin Imager (AMI)}

Sited at the Mullard Radio Astronomy Observatory, Cambridge ($\approx 19\,\rm{m}$ above sea
level), 
AMI consists of a pair of aperture-synthesis interferometric
arrays optimised for SZ-effect imaging centred at 16\,GHz, with six frequency
channels. The Large Array (LA) has a high resolution and flux-density
sensitivity and is used primarily to detect contaminating sources which can
then be subtracted from the Small Array (SA) maps. AMI is described in detail
in \cite{2008MNRAS.391.1545Z} and the technical aspects of the arrays are
summarised in Table \ref{AMI_tech}. The SA has been operating since 2005 (see
e.g. \citealt{2006MNRAS.369.L1}, \citealt{AMI_SA_LDN1111} and
\citealt{AMI_SA_LYNDS}) and the LA since 2008 (see e.g. \citealt{NHW_LA}).
Pointed SZ observations have been straightforward but for blind observations we
have felt it essential to get the best control of systematics that we can -- for
example, we found hard-to-unravel problems with LA pointing and errors in the
electrical lengths of the lags in both the LA and the SA Fourier-transform
correlators that produce small position shifts -- we now have corrections for
these problems that are adequate. We also have very good control over the
influence of radio source contamination (see e.g. \citealt{TMOF_WMAP}, \citealt{FF_MCADAM}, \citealt{LIZ_9C} and \citealt{DAVIES_10C}).

\begin{table}
\caption{AMI technical summary. Note that the brightness sensitivity is highly dependent on the weighting of the visibilities -- in this Table we assume natural weighting.}
 \label{AMI_tech}
\begin{tabular}{lccc}
\hline 
   & SA  & LA   \\\hline 
Antenna diameter & 3.7\,m & 12.8\,m \\ 
Number of antennas & 10 & 8 \\ 
Number of baselines & 45 & 28 \\ 
Baseline length & 5--20\,m & 18--110\,m \\ 
16-GHz power primary beam FWHM  & 19.6$'$ & 5.6$'$ \\ 
Synthesized beam FWHM &  $\approx$ 3$'$ & $\approx$ 30$''$ \\ 
Array flux-density sensitivity & 30\,mJy\,$\rm{s^{-1/2}}$ & 3\,mJy\,$\rm{s^{-1/2}}$ \\ 
Array brightness sensitivity & 4.6\,mK\,$\rm{s^{-1/2}}$ &   16\,mK\,$\rm{s^{-1/2}}$\\
Observing frequency & \multicolumn{2}{|c|}{13.5--18.0\,GHz} \\ 
Bandwidth & \multicolumn{2}{|c|}{3.7\,GHz} \\ 
Number of channels & \multicolumn{2}{|c|}{6} \\ 
Channel bandwidth & \multicolumn{2}{|c|}{0.75\,GHz} \\ 
Polarization measured & \multicolumn{2}{|c|}{I + Q} \\ \hline
 \end{tabular}
\end{table}

\subsection{Observations}

The results presented here are from observations of field AMI002 which is
centred on 02$^{\rm{h}}$\,59$^{\rm{m}}$\,30$^{\rm{s}}$
+26$^\circ$\,16$'$\,30$''$. AMI002 is the first field to have been analysed as
it was the first to reach a target depth of $100 \mu \rm{Jy}$/beam. SA
observations of AMI002 began on 2008 July 19 and ran until 2010 March 3, by
which time 1200 hours of data had been gathered; LA observations  began on 2008
August 8 and ran until 2010 January 10, collecting 630 hours of data. Using
both the SA and the LA the field was typically observed for 8 hours in a day;
this often comprised two individual observations each of 4 hours, split up with
an observation of a flux-density calibrator. Observations were started at
different positions in the field to improve the \textit{uv} coverage.

\begin{figure}
  \centerline{\includegraphics[width=8cm,angle=-90, clip=]{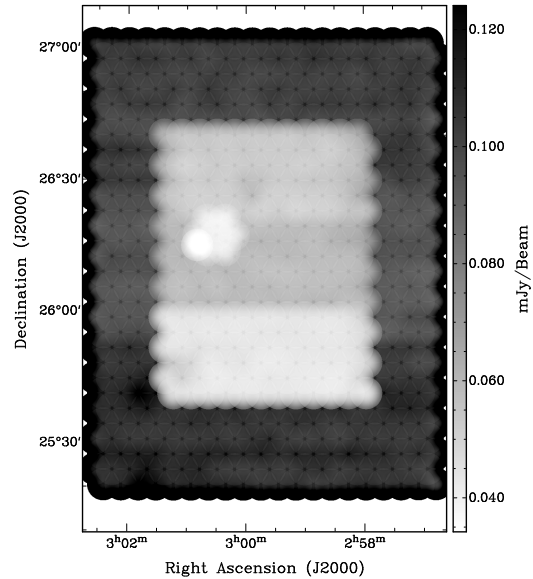}}
  \caption{Noise map for the LA survey field. The inner region noise is $\approx$ $50 \mu \rm{Jy}$, while the noise on the outer region is $\approx$ $100 \mu \rm{Jy}$. The hexagonal region around 03$^{\rm{h}}$~00$^{\rm{m}}$~10$^{\rm{s}}$ +26$^\circ$~15$'$~00$''$ is next to the cluster and was observed to $\approx$ $30 \mu \rm{Jy}$. The inner region of the noise map consists of three subregions; these have slightly different sensitivities due to varying weather conditions and slight differences in observing time\label{SURVEY_LA_NOISE}.}
\end{figure}

\begin{figure}
  \centerline{\includegraphics[width=8cm,clip=,angle=0.]{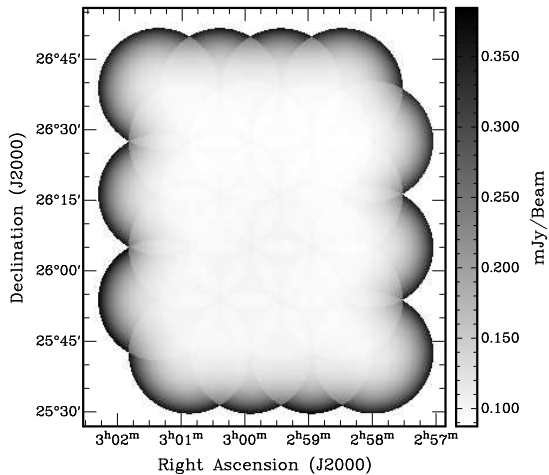}}
  \caption{Noise map for the SA AMI002 field. The noise at the edge of the map increases due to the primary beam of the SA. In the central region the map noise is $\approx$ $100 \mu \rm{Jy}$. This noise map does not include the follow-up SA observations towards 03$^{\rm{h}}$\,00$^{\rm{m}}$\,08.66$^{\rm{s}}$ +26$^\circ$\,15$'$\,16.1$''$  as those data were analysed separately. The noise level of the follow-up SA observations is 65$\mu \rm{Jy}$. \label{SURVEY_SA_NOISE}}
\end{figure}

A rastering technique was used for both the LA and the SA survey observations,
where the pointing centres lie on a 2-D hexagonally-gridded lattice. The LA observations
form a part of the 10C survey data, which are described in detail in
\cite{TMOF_10C}. Additional dedicated pointings towards the cluster candidates
are included to ensure that maximum sensitivity was obtained in the LA maps.
For the 10C survey observations, the pointing centres are separated by 4
arcmin, which allows us to obtain close to uniform sensitivity over the field
while minimising the observing time lost to slewing. In order to detect all
important sources within the SA field, the LA field is slightly larger and the
thermal noise is typically a factor of two lower than the SA thermal noise. To
account for the SA map noise ($\sigma_{\rm{SA,survey}}$) increasing towards the
edge of the field, the LA map consists of two distinct regions, the inner and
the outer. The inner area of the LA field was observed to a noise level of
$\approx$ $50 \mu \rm{Jy}$, whereas the noise in the outer area was
approximately twice as high. The outer region of the LA map is also used to detect bright sources lying just outside the SA field. 
The resulting LA noise map is shown in Figure
\ref{SURVEY_LA_NOISE}. For the SA survey observations the pointing centres are
separated by 13 arcmin giving a close-to-uniform noise level of $\approx$ $100
\mu \rm{Jy}$ over the map. The SA noise map is shown in Figure
\ref{SURVEY_SA_NOISE}. Follow-up SA observations towards the cluster consisted
of 50 hours of data centred at
03$^{\rm{h}}$\,00$^{\rm{m}}$\,08.66$^{\rm{s}}$ +26$^\circ$\,15$'$\,16.1$''$ resulting in a 
noise level of 65$\mu \rm{Jy}$.

The phase calibrator was observed for two minutes every hour using the SA and
for two minutes every ten minutes using the LA. The phase calibrator used for
both the LA and the SA observations was  J0237+2848. The amplitude calibration
for the SA uses 3C286 and 3C48 which are observed daily; the assumed flux
densities are shown in Table \ref{CAL_FLUX_DENSITIES} (\citealt{TMOF_10C}). The LA was
flux-density-calibrated from the SA measurements of J0237+2848; we have adopted
this approach to minimise inter-array calibration errors. Although the flux
density of J0237+2848 is known to vary at AMI frequencies, with a mean
variability index of 3.1 over 269 days (\citealt{TMOF_WMAP}), we ensured that
SA measurements of this source were always within 30 days of the LA
observations. This calibration scheme is described in detail in
\cite{TMOF_10C} and provides flux-density calibration errors of less than 5\%.

\begin{table}
 \caption{Assumed flux densities for the SA flux-density calibrators.}
 \label{CAL_FLUX_DENSITIES}
\begin{tabular}{lcccc}
\hline 
Channel  & $\rm{\bar{\nu}/GHz}$ & \multicolumn{2}{|c|}{$\rm{S_{I+Q}/Jy}$}  \\
         &        & 3C48  &  3C286 \\  \hline
1        & 14.2   & 1.850 &  3.663  \\ 
2        & 15.0   & 1.749 &  3.535 \\ 
3        & 15.7   & 1.658 &  3.414  \\ 
4        & 16.4   & 1.575 &  3.308  \\ 
5        & 17.1   & 1.500 &  3.206  \\ 
6        & 17.9   & 1.431 &  3.111  \\ \hline
 \end{tabular}
\end{table}

\subsection{Data Reduction}

There are 65 LA observations and 337 SA observations of AMI002, each being
passed through \textsc{reduce}, the in-house software developed for the VSA and
AMI data reduction. \textsc{reduce} was used to flag telescope pointing errors,
shadowing effects and hardware errors.  The data are also flagged for
interference before being Fourier transformed into the frequency domain, where
they are corrected for system-temperature variations,  phase-calibrated and
amplitude-calibrated. In the frequency domain the data are again searched for
interference and baselines with inconsistent flux-density values are flagged.
The data are reweighted so that baselines and channels with the lowest noise
have the highest weight. The data are then stored as \textsc{uvfits} files, with each raster pointing being treated as an independent source within the \textsc{fits} definition.
This reduction scheme follows that of \cite{DAVIES_09}. Individual
\textsc{uvfits} files for the LA and the SA are combined into a single
multisource \textsc{uvfits} file for each array, which are taken into
\textsc{aips} \footnote{http://www.aips.nrao.edu} for imaging.

\begin{figure*}
\centerline{\includegraphics[width=7.5cm,height= 7.5cm,clip=,angle=0.]{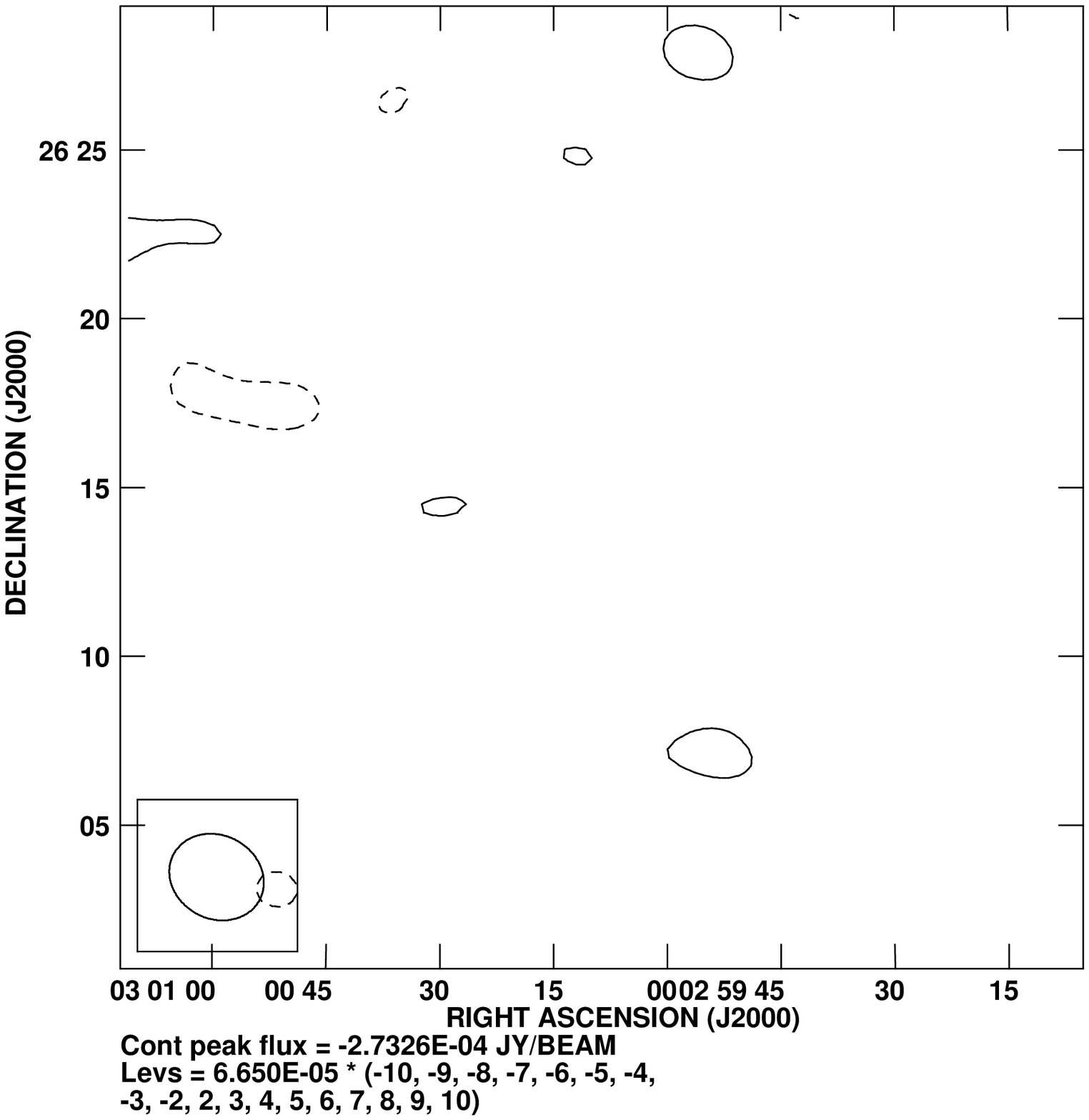}\qquad\includegraphics[width=7.5cm,height= 7.5cm,clip=,angle=0.]{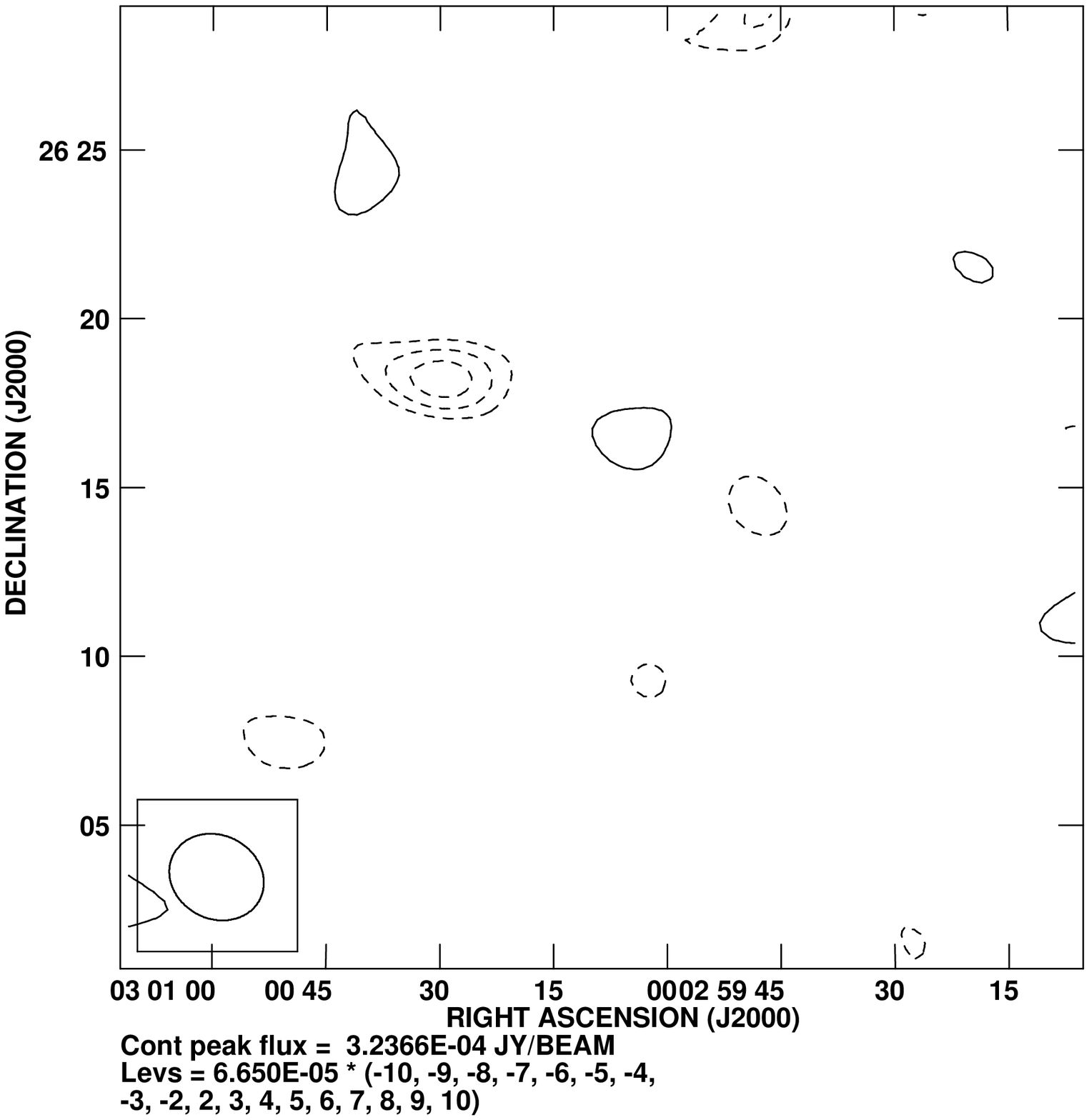}}
\caption{The jack-knifed SA pointed observations of the cluster candidate (an image of the data is shown on the left of Figure \ref{fig:SA_pointed}). On the left, data are split into plus and minus baseline, one the right, data are split according to median date. The contour levels are linear from 2 $\sigma_{SA,survey}$ to 10 $\sigma_{SA,survey}$ ($\sigma_{SA,pointed}$ $=$ 65$\mu \rm{Jy}$); positive contours are solid lines and negative contours are dashed lines. The ellipse at the bottom left of the maps shows the SA synthesised beam.
\label{fig:SA-pointed-jackknife}}
\end{figure*}

The SA data were checked for systematics using two jack-knife tests. In test (a), calibrated visibilities from ``plus'' correlator boards are subtracted from those obtained from ``minus'' correlator boards -- the signal is the same for both correlations but the latter inserts an additional 180$^{\circ}$ phase shift into the signal from one antenna (see \citealt{Holler_2007} for a full description of the AMI correlator). For test (b), data obtained before the weighted median date of the visibilities are subtracted from data obtained afterwards. Either test will remove signals present in both halves of the data but noise or systematics that vary with time will remain. For the follow-up pointed SA observations presented in this paper test (a) revealed no systematics and test (b) showed a negative feature with a flux-density of 0.35mJy/beam associated with the 2.26mJy/beam source at 03:00:29.46 +26:18:39.9. Investigation demonstrated that this residual was a consequence of the flux-density of the source being dependent upon the elongation of the synthesized beam. For reasons of scheduling, it became clear that the synthesized beam from the first half of data was elongated in the NW-SE direction and the source was measured to have a flux-density of 2.51mJy/beam. In the second half of the data the synthesized beam was extended in the NE-SW direction and the measured flux-density was 2.03mJy/beam. Maps of the jack-knifed pointed SA observations are shown in Figure \ref{fig:SA-pointed-jackknife}.

\subsection{LA map-making and source-finding}

LA maps for each AMI channel and the continuum were produced for each of the
pointings within the AMI002 field using the \textsc{aips} task \textsc{imagr}.
The maps are \textsc{clean}ed to three times the map thermal-noise without any
individual \textsc{clean} boxes. The individual pointings are combined using
the \textsc{flatn} task, discarding data lying outside the 10\% point of the power
primary beam. \textsc{flatn} is also used to create appropriately-weighted
noise maps using the thermal noise levels in the individual pointings.

Source finding is carried out using the LA continuum map with the AMI
\textsc{sourcefind} software. All pixels on the map with a flux density greater
than $0.6 \times 4 \times \sigma_n$, where $\sigma_n$ is the noise map value
for that pixel, are identified as peaks. The flux densities and positions of
the peaks are determined using a tabulated Gaussian sinc degridding function to
interpolate between the pixels. Only peaks where the interpolated flux density
is greater than $4 \times \sigma_n$ are identified as sources. The
\textsc{aips} routine \textsc{jmfit} fits a two-dimensional Gaussian to each
source to give the angular size and the integrated flux density for the source.
These fitted values are compared to the point-source response function of the
telescope to determine whether the source is extended on the LA map. The
mapping and the source finding techniques are described in more detail in
\cite{TMOF_10C}.

For each source we use the \textsc{sourcefind} algorithm to find the flux
densities in the individual AMI LA channel maps at the positions of the
detected sources. By assuming a power-law relationship between flux density and
frequency ($S \propto \nu^{-\alpha}$) we use the channel flux-densities to
determine the spectral index $\alpha$ for each source. The spectral index is
calculated using an MCMC method based on that of \cite{Metromod} -- the prior
on the spectral index has a Gaussian distribution with a mean of 0.5 and
$\sigma$ of 2.0, truncated at $\pm 5.0$. The minimum spectral index of a source
in the AMI002 field was found to be 0.0 and the maximum was 1.8. The map noise
in each channel map at the position of the source was used to calculate the weighted mean of the channel frequencies and  determine the effective central frequency $\nu_{0}$ of the 
source. The effective central frequency varies
between pointings due to flagging applied in \textsc{reduce}.  Unlike in
\cite{TMOF_10C}, the data are not reweighted to the same frequency because this
leads to a small loss of sensitivity.

In total we detect 203 sources in the AMI002 LA map at four times the LA map 
noise ($\sigma_{\rm{LA}}$), 11 of which are extended. The most extended source 
has an area of 1.9 LA synthesized beams. As the SA synthesized beam is significantly
larger we do not expect any extended sources in the SA map. For each source we
catalogue the right-ascension $\rm{x_{s}}$, declination $\rm{y_{s}}$, flux
density at the central frequency $S_{0}$, spectral index and the central
frequency. If a source is extended we use the centroid of the fitted Gaussian
as the position and the integrated flux density instead of the peak flux
density.

\section{Identifying and modelling cluster candidates}\label{sec:METHOD}

Our analysis necessarily depends in part on the fact that we do not know -- in the absence of e.g. optical spectroscopic observations -- the redshift of the blind SZ clusters. We have thus carried out our  analysis in two main ways, both fully Bayesian and based on \citealt{MH_MCADAM}, \citealt{MARSH_MCADAM} and \citealt{FF_MCADAM}, as follows. 

(1) Physical model. We assume an isothermal $\beta$-profile for the gas density as a function of radius; we assume all the cluster kinetic energy is in the internal energy of the cluster gas and that the relation between gas temperature and total cluster mass is then given by the virial theorem; and we assume the prior probability for the comoving number density of clusters as a function of total mass and redshift is given by previous theoretical/simulation work -- we here use the predictions of  \cite{Evrard_02} and \cite{Jenkins} and note that more recent such work does not make a substantial difference for our purposes. With these assumptions we are then able to (a) estimate the significance of an SZ detection, and (b) produce probability distributions of physical cluster parameters such as mass and radius. For both (a) and (b) the methodology takes into account radio sources, receiver noise, and the statistical properties of the primordial CMB structure; it cannot take into account other effects that have not been dealt with in, for example, telescope design, telescope commissioning, observing and data reduction.

(2) Phenomenological model. Some or all of the assumptions in (1) may be poor or wholly wrong. Accordingly in (2), we make far fewer assumptions. We assume isothermality and that the temperature decrement as a function of angular distance is given by a $\beta$-model. This model cannot give probability distributions of values of physical importance such as mass, but still does give the significance of the SZ detection in the presence of radio sources, receiver noise and primordial CMB structure; like (1) it cannot take into account other effects that have not previously been dealt with.

We give the significance of decrement detection in a third way, the decrement signal in units of receiver noise. We point out that for AMI this method of course takes no account of primordial CMB structures but does take into account radio sources, and the higher flux-density sources have had their SA flux-densities estimated in a Bayesian way from SA data and priors from LA measurements -- this allows for inter-array calibration problems and for LA and SA observations that were not precisely simultaneous. Again, like (1) and (2), this method cannot take into account other effects that have not previously been dealt with.

\subsection{Physical model}

Our primary Bayesian analysis is based on a physical model for the
cluster producing the SZ effect.  The SA observations of the AMI002
survey field are analysed using a model characterised by the
parameters $\Theta = (\Theta_{c}, \Psi)$, where $\Theta_{c} =
(x_{c},y_{c},\phi,f, \beta, r_{c}, M_{T,200}, z)$ are cluster
parameters and $\Psi = (x_{s},y_{s},S_{0},\alpha)$ are source
parameters (\citealt{FF_MCADAM}). Here $x_{c}$ and $y_{c}$ give the
cluster position, $\phi$ is the orientation angle measured from N
through E, $f$ is the ratio of the lengths of the semi-minor to
semi-major axes, $\beta$ describes the cluster gas density $\rho_{g}$
according to \cite{BETA_1}, where the gas density decreases with
radius $r$
\begin{equation}
 \rho_{g}(r) =\frac{\rho_{g}(0)}{[1+(r/r_{c})^{2}]^{\frac{3\beta}{2}}},
 \label{eqn:Beta_profile}
\end{equation}
$r_{c}$ is the core radius, $M_{T,200}$ is the cluster total mass
within a radius $r_{200}$ and $z$ is the cluster redshift. $r_{200}$
is defined as the radius inside which the mean total density is 200
times the critical density $\rm{\rho_{crit}}$. \cite{FF_MCADAM} and
\cite{CARMEN} describe the parameters and the methods used to extract
these from the data in more detail.  For this work we sample from
$x_{c}$, $y_{c}$, $\phi$,$f$, $\beta$, $r_{c}$, $M_{T,200}$ and $z$
and derive other cluster parameters such as the cluster gas mass
$M_{g,200}$ and the cluster temperature $T$. We also assume a
mass-temperature relationship characteristic of a virialised cluster;
this is the favoured model (M3) in \cite{CARMEN}, although we sample
from $M_{T,200}$ rather than $M_{g,200}$, see also \citealt{Olamaie_2010}. 
The total cluster mass within $r_{200}$ is
\begin{equation}
M_{T,200} = \frac{4 \pi}{3} r^{3}_{200} (200 \rho_{crit}). 
\end{equation}
The gas fraction $f_{g}$ is derived from the results of
\cite{WMAP_FGAS_VALUE} taking into account our value for $h$ and that
the gas-mass fraction is $\approx 0.9$ of the baryonic mass fraction.
The ellipticity of the clusters is calculated by applying a coordinate
transformation from point ($\theta_{1},\theta_{2}$) on the sky:
\begin{equation}
 \left( \begin{array}{c}
 \theta_{1}' \\
 \theta_{2}' \end{array} \right) =
 \left( \begin{array}{cc}
 \sqrt{f} &  0 \\
 0   &  1/\sqrt{f} \end{array} \right)
 \left( \begin{array}{cc}
 \cos \phi   & \sin\phi \\
 -\sin \phi  & \cos \phi \end{array} \right)
 \left( \begin{array}{c}
 \theta_{1} \\
 \theta_{2} \end{array} \right).
\end{equation}
%
%
Lines of constant $\theta'$ represent ellipses enclosing an area $\pi
a b$, where $a$ is the semi-minor axis and $b$ is the semi-major
axis. This transforms the circular slices perpendicular to the line of
sight to an ellipse, keeping the area of the ellipse the same as the
circular slice.  A summary of the priors used on the model parameters
is shown in Table \ref{MC_TRI_PRIORS}.

\begin{table*}
\caption{Priors used for the Bayesian analysis assuming a physical
  cluster model.}
 \label{MC_TRI_PRIORS}
\begin{tabular}{lcc}
 \hline
Parameter & Prior \\  \hline
Source position ($x_{s}$) & A delta-function prior using the LA positions \\ 
Source flux density ($S_{0}/\rm{Jy}$) & A Gaussian centred on the LA continuum value with a $\sigma$ of 40\% \\ 
Source spectral index ($\alpha$) & A Gaussian centred on the value calculated from the LA channel maps with the LA error as $\sigma$  \\ 
Redshift ($z$) & Joint prior with $M_{T}$ between 0.2 and 2.0 (\citealt{Jenkins} or \citealt{Evrard_02})\\ 
Core radius ($r_{c}/h_{70}^{-1}\rm{kpc}$) & Uniform between 10 and 1000 \\ 
Beta   ($\beta$)  & Uniform between 0.3 and 2.5 \\ 
Mass ($M_{T,200}/h_{70}^{-1}M_{\odot}$) & Joint prior with $z$ between 2.0 $\times$ $10^{14}$ and 5 $\times$ $10^{15}$ (\citealt{Jenkins} or \citealt{Evrard_02})\\ 
Gas fraction ($f_{g}/h_{70}^{-1}$) & Delta-function prior at 0.11 (\citealt{WMAP_FGAS_VALUE}) \\ 
Cluster Position ($\bf x_{c}$) & Uniform search triangle (Figure \ref{SA_triangle}) \\ 
Orientation angle ($\phi/\deg$) & Uniform between 0 and 180 \\ 
Ratio of the length of semi-minor to semi-major axes ($f$) & Uniform between 0.5 and 1.0 \\ \hline
\end{tabular}
\end{table*}

The above approach has already been used to detect the SZ effect from
AMI observations of known clusters in \cite{7CLUSTERS} and
\cite{CARMEN}. However, for blind cluster surveys we are faced with
the additional problem that we do not have a priori evidence for a
cluster at a particular position (or redshift). In analysing a survey
field, the marginalised posterior distribution in the
$(x_c,y_c)$-plane will typically contain a number of local peaks; some
of these may correspond to the presence of a real cluster, whereas
others may result from chance statistical fluctuations in the
primordial CMB and/or instrument noise. Each local peak in the
posterior is automatically identified by the {\sc MultiNest} sampler
(\citealt{MULTINEST_1} and \citealt{MULTINEST_2}) used in our Bayesian analysis, and may subsequently be analysed independently to obtain cluster parameter estimates.

To determine the significance of each such putative cluster detection,
we perform a Bayesian model selection, which makes use of estimated
cluster number counts from analytical theory (e.g. the
\citealt{Evrard_02} approximation to \citealt{Press_Sch}) and
numerical modelling (e.g. \citealt{Jenkins}) together with
measurements of the rms mass fluctuation amplitude on scales of size
8~${h_{100}^{-1}}$Mpc at the current epoch (see
e.g. \citealt{2DF_SIGMA8}, \citealt{SDSS_SIGMA8} and
\citealt{CHANDRA_SIGMA8}). It must be borne in mind, however, that the
actual values of the number density of clusters, particularly at high
redshift, are uncertain and hence the degree of applicability of these
as priors is unclear.

In our Bayesian model selection, we calculate the formal Bayesian probability of two
hypotheses: the first, $\Pr(H_{\geq1} |D)$, assumes at least one
cluster with $M_{T,{\rm lim}} < M_{T,200} < M_{T,{\rm max}}$ is
associated with the local peak in the posterior distribution under
consideration; the second, $\Pr(H_{0} |D)$, assumes no such cluster is
present. Here $M_{T,{\rm lim}}$ is the limiting cluster mass that can
be detected and $M_{T,{\rm max}}$ is the maximum mass of a cluster. In
particular, we consider the ratio $R$ (also known as the Bayes factor, or the odds)
of these two formal probabilities
\begin{equation}
R \equiv \frac{\Pr(H_{\geq1} |D)}{\Pr(H_{0}|D)}.
\label{eqn:R}
\end{equation}
To evaluate this ratio, let us first denote by $S$ the area
in the $(x_c,y_c)$-plane of the `footprint' of the local posterior peak
under consideration (we will see below that a precise value for $S$ is
not required). Also, we denote by $H_{n}$ the hypothesis that there are
$n$ clusters with $M_{T,{\rm lim}} < M_{T,200} < M_{T,{\rm max}}$ with
centres lying in the footprint $S$, so that
\begin{equation}
  \Pr(H_{\geq1}) = \sum ^{\infty}_{n=1} \Pr(H_n).
\end{equation}
Thus equation (\ref{eqn:R}) can be written as
\begin{equation}
R = \frac{\sum ^{\infty}_{n=1}  \Pr(H_{n}|D)}{\Pr(H_{0}|D)}
= \frac{\sum ^{\infty}_{n=1} \Pr(D|H_{n})\Pr(H_{n})}
{\Pr(D|H_{0})\Pr(H_{0})},
\label{eqn:R3}
\label{eqn:R2}
\end{equation}
where we have used Bayes' theorem in the second equality.
Assuming that objects are randomly distributed over the sky, then
\begin{equation}
\Pr(H_{n}) =\frac{e^{-\mu_{S}} \mu_{S}^{n}}{n!} 
\label{eqn:prob}
\end{equation}
where  $\mu_S$ is the expected number of clusters with  
$M_{T,{\rm lim}} < M_{T,200} < M_{T,{\rm max}}$  in a region $S$.
This is given by $\mu_S=S\mu$, where $\mu$ is the expected number of
clusters per unit sky area:
\begin{equation}
\mu = \int^{z_{{\rm max}}}_{z_{{\rm min}}} \int ^{M_{T,{\rm max
}}}_{M_{T,{\rm lim}}} \frac {d^2n}{dMdz} dMdz,
\label{eqn:mu}
\end{equation}
where $n(z,M)$ is the comoving number density of clusters as a
function of redshift and mass. For the calculation of $\mu$, we
follow the method of either \cite{Evrard_02} or
\cite{Jenkins}. If we further assume that
there is very low probability of two or more clusters having their centres
in the region $S$ ($\mu_S\ll 1$) we can neglect $\mu_{S}^2$ and larger 
powers of $\mu_S$, so that equation
(\ref{eqn:R3}) can be approximated simply by
\begin{equation}
R \approx \frac{Z_{1}(S)\mu_S}{Z_{0}}.
\label{eqn:R5}
\end{equation}
where the $Z_{1}(S) = \Pr(D|H_{1})$ is the `local evidence' (see
Feroz et al. 2009) associated with the posterior peak under
consideration in the single-cluster model, and  
$Z_{0} = \Pr(D|H_{0})$ is the `null'evidence (which does
not depend on $S$). 

Our Bayesian analysis uses \textsc{MultiNest} to calculate the
Bayesian evidence for the different hypotheses (\citealt{MULTINEST_1}
and \citealt{MULTINEST_2}).  When searching for clusters in some
survey area $A$, however, a uniform prior $\pi(x_c,y_c)=1/A$ is
assumed on the position of any cluster, rather than assuming a uniform
prior over the footprint $S$.  Thus, \textsc{MultiNest} returns a
local evidence associated with the posterior peak that is given by
\begin{equation}
\tilde{Z}_1(S)  = \frac{S}{A} Z_1(S),
\end{equation}
and the `null' evidence $\tilde{Z}_0 = Z_0$ remains unchanged. Thus,
if we denote the expected number of clusters in the survey area by
$\mu_A = (A/S) \mu_S$, then (\ref{eqn:R3}) becomes
\begin{equation}
R \approx \frac{\tilde{Z}_{1}(S)\mu_A}{\tilde{Z}_{0}}.
\label{eqn:R5}
\end{equation}
Here $\tilde{Z}_{1}(S)$ and $\tilde{Z}_{0}$ are outputs of {\sc
  MultiNest} and $\mu_A$ is easily calculated from (\ref{eqn:mu})
given some assumed cluster mass function, and so $R$ may then be
calculated. In our analysis $\mu_A < 1$ and the $R$ value that we
calculate is smaller than that obtained by setting the prior ratio equal to unity.
\cite{Jeffreys_1961} provides an interpretive scale for
the $R$ value, as do revised scales  such as \cite{Gordon_2007}. 
Moreover, the $R$ value in (\ref{eqn:R5}) can be turned
into a formal Bayesian probability $p$ that the putative detection is indeed due to a
cluster with mass $ M_{T,{\rm lim}} < M_{T,200} < M_{T,{\rm max}}$ and
centre lying in $S$, which is given by
\begin{equation}
 p = \frac{R}{1+R}.
 \label{eqn:P1}
\end{equation}

\subsection{Phenomenological model}

An alternative approach is to set aside the physical cluster model and
instead adopt a model based on a phenomenological description of the
SZ decrement itself.

In this case, at the location of each putative cluster detection
identified using the physical cluster model, we simply fit a $\beta$
profile to the SZ temperature decrement using the parameters
$\theta_{c}$, $\beta$ and $\Delta \rm{T_{0}}$ to characterise shape
and magnitude of the decrement according to
\begin{equation}
\Delta T_{SZ} = \Delta T_{0} \left(1+ \frac{\theta^{2}} {\theta_{c}^{2}} \right) ^ {(1- \frac{3\beta}{2})}.
\label{eqn:blob_profile}
\end{equation}
The assumed priors on these parameters are summarised in Table
\ref{tab:blob_priors}. 
\begin{table}
\caption{Priors used for the Bayesian analysis assuming a phenomenological cluster model.}
 \label{tab:blob_priors}
\begin{tabular}{lcc}
 \hline
Parameter                & Prior   \\  \hline
$\Delta T_{0}$           &  Uniform between $\pm3000\mu \rm{K}$ \\ 
$\theta_{c}$             &  Uniform between 20$''$ and 500$''$\\ 
$\beta$                  &  Uniform between 0.4 and 2.5 \\ \hline
\end{tabular}
\end{table}
In this analysis
we continue to use Gaussian priors on the flux densities and on the
spectral indices of significant sources, and delta-function priors for
faint sources. We also assume a Gaussian prior ($\sigma=60''$) on position
centred on each decrement.

This approach allows us to produce a posterior distribution that
directly describes the temperature decrement and also allows us to
evaluate what proportion of the decrement is caused by the SZ effect,
while also accurately accounting for point sources, receiver noise and
the statistical properties of the primary CMB anisotropy. 

\section{The analysis}\label{sec:BAYES}

\begin{figure}
\centerline{\includegraphics[clip=true,width=8.cm,angle=-90]{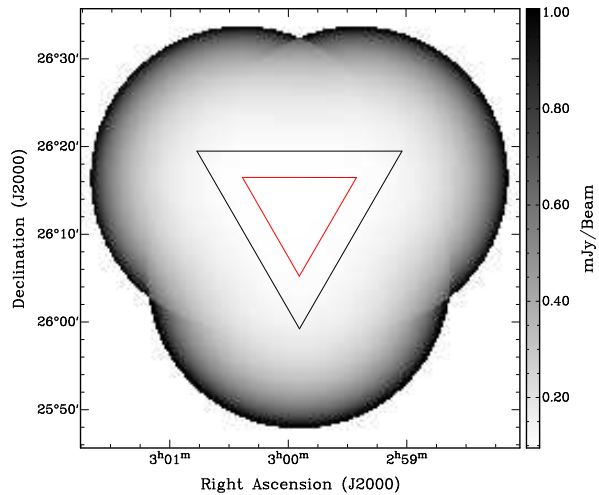}}
\caption{Noise map for a SA triangle of observations out to the 0.1 contour of the power primary beam. The inner triangle is between the pointing centres; the outer triangle is the area that is searched for clusters with our Bayesian analysis.\label{SA_triangle}}
\end{figure}

The AMI002 SA survey map contains 24 individual pointing centres. A
single Bayesian analysis of the entire field is prohibitively
computationally expensive because of the large quantity of data and
the high dimensionality of the parameter space. Instead three
pointings are analysed at a time. Each set of three pointing centres
form a triangle and in total there are 30 different triangles in the
AMI002 field, an example of which is shown in Figure \ref{SA_triangle}.

To reduce the dimensionality of the parameter space further, all
sources located at positions where the primary beam has fallen below 
10$\%$ of its maximum, together
with sources that have a flux density measured on the LA that is
lower than $4\sigma_{\rm{SA,survey}}$, are given delta-function priors
on their positions, spectral indices and flux densities. We search for
clusters in a triangular area which is an enlarged version of the
triangle formed between the pointing centres -- the radius of the
inscribed circle is 3$'$ larger. This allows us to detect clusters
out to the edge of our most sensitive areas and ensures that the
search areas belonging to adjacent triangles overlap. The minimum rms
noise within a search triangle in the AMI002 field is $\approx$
$\rm{100\mu Jy}$ and the maximum is $\approx$ $\rm{140\mu Jy}$.  The
limiting cluster total mass is set to $\rm{M_{T,{\rm lim}} = 2 \times
  10^{14} h_{70}^{-1}M_{\odot}}$ and the maximum cluster mass to $\rm{M_{T,{\rm
      max}} = 5 \times 10^{15} h_{70}^{-1}M_{\odot}}$. The limiting mass is
conservative given the radio flux-density sensitivity of our
observations.

We follow up our most significant detections with pointed observations towards
the candidate. The data from these observations can be analysed with our
Bayesian method with lower dimensionality because there are fewer sources
within 0.1 of the power primary beam with flux densities greater than
$4\sigma_{\rm{SA,pointed}}$.
For the follow-up pointed observations the prior on the cluster position is
altered to a 1000$''$ x 1000$''$ box centred on the pointing centre and we
allow our Bayesian analysis software to fit the source positions with a
Gaussian prior centred on the LA position with an error of 5$''$.

\section{Results and Discussion}\label{sec:RESULTS}

The most significant candidate cluster detection made using the
Bayesian analysis of the AMI SA survey field AMI002 is located at
J~03$^{\rm{h}}$~00$^{\rm{m}}$~16.5$^{\rm{s}}$
+26$^\circ$~13$'$~59.5$''$, where {\sc MultiNest} identifies a single
marginalised posterior peak in the $(x_c,y_c)$-plane centred on this
location. The significance of the cluster detection is $R \approx 8.7$
when we use Model (1) and the \cite{Evrard_02} prior and $R \approx 26$ when we
apply Model (1) using the \cite{Jenkins} prior.
The relevant area of the survey field is shown in Figure
\ref{fig:TRI_sig2noise} before and after source subtraction (see
below).  In the search triangle that contains our cluster candidate
there are 59 sources within 0.1 of the power primary beam, 43 of which
have a flux density below $4\sigma_{\rm{SA,survey}}$; the other 16
have been modelled with our Bayesian analysis.  The location of the
marginalised posterior peak is indicated by the small box in the
figures.

\begin{figure*}
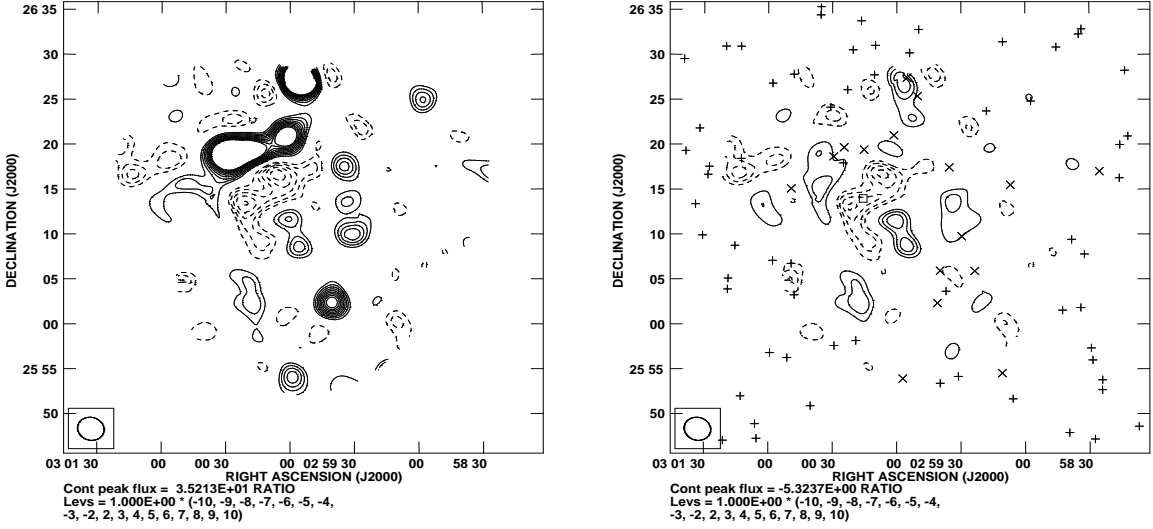

\centerline{\includegraphics[width=7.5cm,height= 7.5cm,clip=,angle=0.]{TRI_sig2noise_a.ps}\qquad\includegraphics[width=7.5cm,height= 7.5cm,clip=,angle=0.]{TRI_sig2noise_a_sub_P11_14_15B.ps}}
\caption{The SA survey-field observations of the cluster candidate. On the left is the map before source subtraction and on the right is after the map after the sources in Table \ref{AMI002_sources} have been subtracted. The + symbols indicate the positions of sources with flux densities less than 4 $\sigma_{\rm{SA,survey}}$ and the $\times$ symbols represent sources which have a flux density greater then 4 $\sigma_{\rm{SA,survey}}$. The box on the source-subtracted map shows the position of the cluster candidate. These maps are signal-to-noise maps. Each pointing has a thermal-noise level of approximately 0.11~mJy. The contour levels are linear from 2 $\sigma_{SA,survey}$ to 10 $\sigma_{SA,survey}$; positive contours are solid lines and negative contours are dashed lines. The ellipse at the bottom left of the maps shows the SA synthesised beam. \label{fig:TRI_sig2noise}}
\end{figure*}

At this position in our survey field is a highly-extended,
non-circular negative feature with a peak flux-density decrement of
$\approx$ 0.6~mJy ($5\sigma_{\rm{SA,survey}}$). SA observations are
mapped in \textsc{aips} using the same method as for the LA, but with
a pixel size of $15\,\rm{arcsec}$. We subtract sources from the
\textit{uv}-\textsc{fits} data using the in-house software
\textsc{muesli}. \textsc{muesli} performs the same function as the
\textsc{aips} task \textsc{uvsub}; however, it is optimised for
processing AMI data. 
The parameters of the 16 modelled sources are shown in Table
\ref{AMI002_sources}; we find no evidence that any of them is extended relative
to the LA synthesised beam. The source subtraction leaves very little residual
flux density on the map, indicating that the phase stability and calibration of
AMI is robust. The most significant source-subtraction residuals are towards
the edge of the SA power primary beam where we expect the phase errors to be
larger and the beam model to be less accurate.

\begin{table*}
\caption{The fitted parameters for the 16 sources with LA flux densities greater than 4 $\sigma_{\rm{SA,survey}}$ ($65 \mu Jy$). This includes the nine sources with flux densities greater than $4\sigma_{\rm{SA,pointed}}$. The positions and mean frequencies are from LA observations, whereas the flux densities and spectral indices are the values obtained from our Bayesian analysis of the SA survey field.}
 \label{AMI002_sources}
\begin{tabular}{lccccc}
 \hline
Right ascension & Declination  & Flux density    & Spectral index & Mean frequency & Flux density $ > 4\sigma_{\rm{SA,pointed}}$ \\
(J2000)         & (J2000)      & (mJy)    &                & (\,GHz)          &  \\  \hline

03:00:24.53 &  +26:19:40.83 & 1.21 $\pm$ 0.12 &    +1.38 $\pm$ 0.38 & 15.63 & $\surd$ \\ 
03:00:29.46 &  +26:18:39.95 & 2.26 $\pm$ 0.12 &    +0.71 $\pm$ 0.29 & 15.64 & $\surd$ \\ 
02:59:06.92 &  +26:15:29.59 & 0.26 $\pm$ 0.09 &    +1.46 $\pm$ 1.10 & 15.57  & $\surd$ \\ 
02:59:50.35 &  +26:25:22.37 & 0.23 $\pm$ 0.11 &    +0.51 $\pm$ 1.24 & 15.58 & $\times$ \\ 
02:59:39.76 &  +26:05:56.15 & 0.40 $\pm$ 0.09 &    +2.31 $\pm$ 1.06 & 15.52   & $\times$\\ 
03:00:15.23 &  +26:19:25.56 & 1.44 $\pm$ 0.11 &    +1.59 $\pm$ 0.40 & 15.64  & $\surd$\\ 
02:59:23.57 &  +26:05:54.53 & 0.40 $\pm$ 0.10 &    +0.81 $\pm$ 1.24 & 15.54  & $\times$\\ 
02:59:55.16 &  +26:27:26.24 & 8.49 $\pm$ 0.22 &    +0.33 $\pm$ 0.07 & 15.59  & $\surd$\\ 
02:59:10.71 &  +25:54:31.60 & 3.81 $\pm$ 0.41 &    +1.05 $\pm$ 0.16 & 15.57  & $\times$\\ 
02:59:35.43 &  +26:17:26.77 & 0.64 $\pm$ 0.09 & $-$0.22 $\pm$ 0.94 & 15.53  & $\surd$\\ 
03:00:49.28 &  +26:15:05.70 & 0.53 $\pm$ 0.12 &    +0.40 $\pm$ 0.42 & 15.67  & $\surd$\\ 
02:59:29.68 &  +26:09:46.99 & 0.50 $\pm$ 0.07 &    +1.71 $\pm$ 1.26 & 15.55  & $\times$\\ 
02:59:41.05 &  +26:02:20.41 & 1.58 $\pm$ 0.12 &    +1.36 $\pm$ 0.36 & 15.54 & $\surd$ \\ 
02:59:57.17 &  +25:53:56.17 & 1.40 $\pm$ 0.23 &    +2.20 $\pm$ 0.34 & 15.58  & $\times$\\ 
03:00:01.33 &  +26:21:01.55 & 1.96 $\pm$ 0.12 & $-$0.45 $\pm$ 0.31 & 15.56 & $\surd$ \\ 
02:58:25.32 &  +26:16:59.59 & 1.63 $\pm$ 0.33 &    +0.95 $\pm$ 0.27 & 15.58  & $\times$\\ \hline

 \end{tabular}
\end{table*}

\begin{figure*}
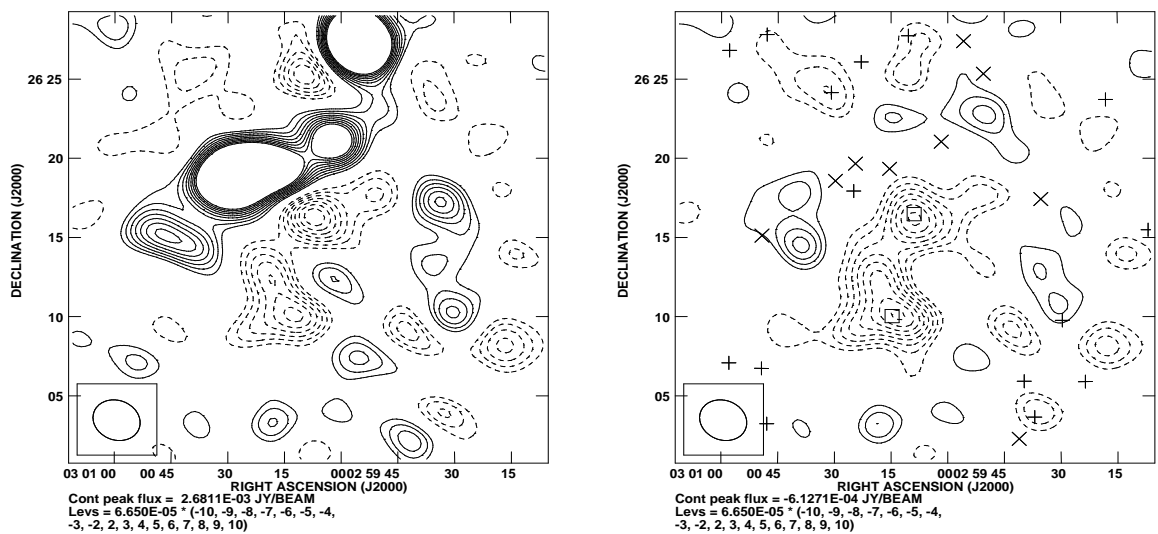

\centerline{\includegraphics[width=7.5cm,height= 7.5cm,clip=,angle=0.]{SA_pointed_a.ps}\qquad\includegraphics[width=7.5cm,height= 7.5cm,clip=,angle=0.]{SA_pointed_a_sub_T323M_box.ps}}
\caption{The AMI SA pointed observation towards the cluster. On the left is the map before source-subtraction and on the right is the map after the relevant sources in Table \ref{AMI002_sources} have been subtracted. The + symbols indicate the positions of sources with flux densities less than 4 $\sigma_{\rm{SA,pointed}}$, the $\times$ symbols represent sources which have a flux density greater than 4 $\sigma_{\rm{SA,pointed}}$ on the SA map. For this run, the sampler has been allowed to fit the positions of the $\times$ type sources with a Gaussian centred on the LA source position. The contour levels are linear from 2 $\sigma_{SA,pointed}$ to 10 $\sigma_{SA,pointed}$ ($\sigma_{SA,pointed}$ $=$ 65$\mu \rm{Jy}$); positive contours are solid lines and negative contours are dashed lines. The boxes indicate the positions of the cluster candidates. Candidate 1 is at J~03$^{\rm{h}}$~00$^{\rm{m}}$~14.8$^{\rm{s}}$ +26$^\circ$~10$'$~02.6$''$ and candidate 2 is at J~03$^{\rm{h}}$~00$^{\rm{m}}$~08.9$^{\rm{s}}$ +26$^\circ$~16$'$~29.1$''$. When imaging the source subtracted map \textsc{clean} boxes have been placed around each candidate. The ellipse at the bottom left of the maps shows the SA synthesised beam.
\label{fig:SA_pointed}}
\end{figure*}

\begin{figure*}
\centerline{\includegraphics[width=5.0cm,height= 5.0cm,clip=,angle=0.]{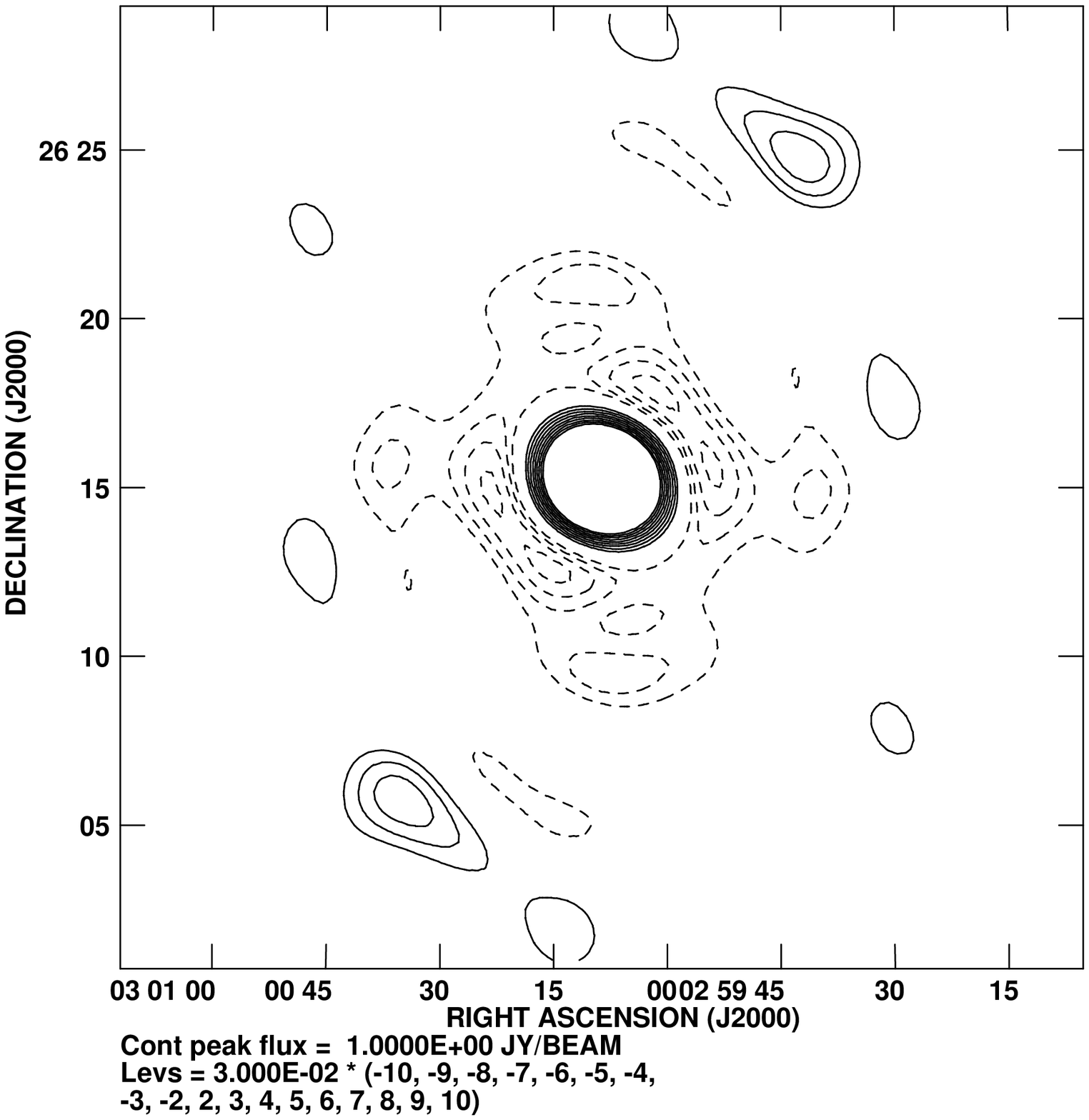}\qquad\includegraphics[width=5.0cm,height= 5.0cm,clip=,angle=0.]{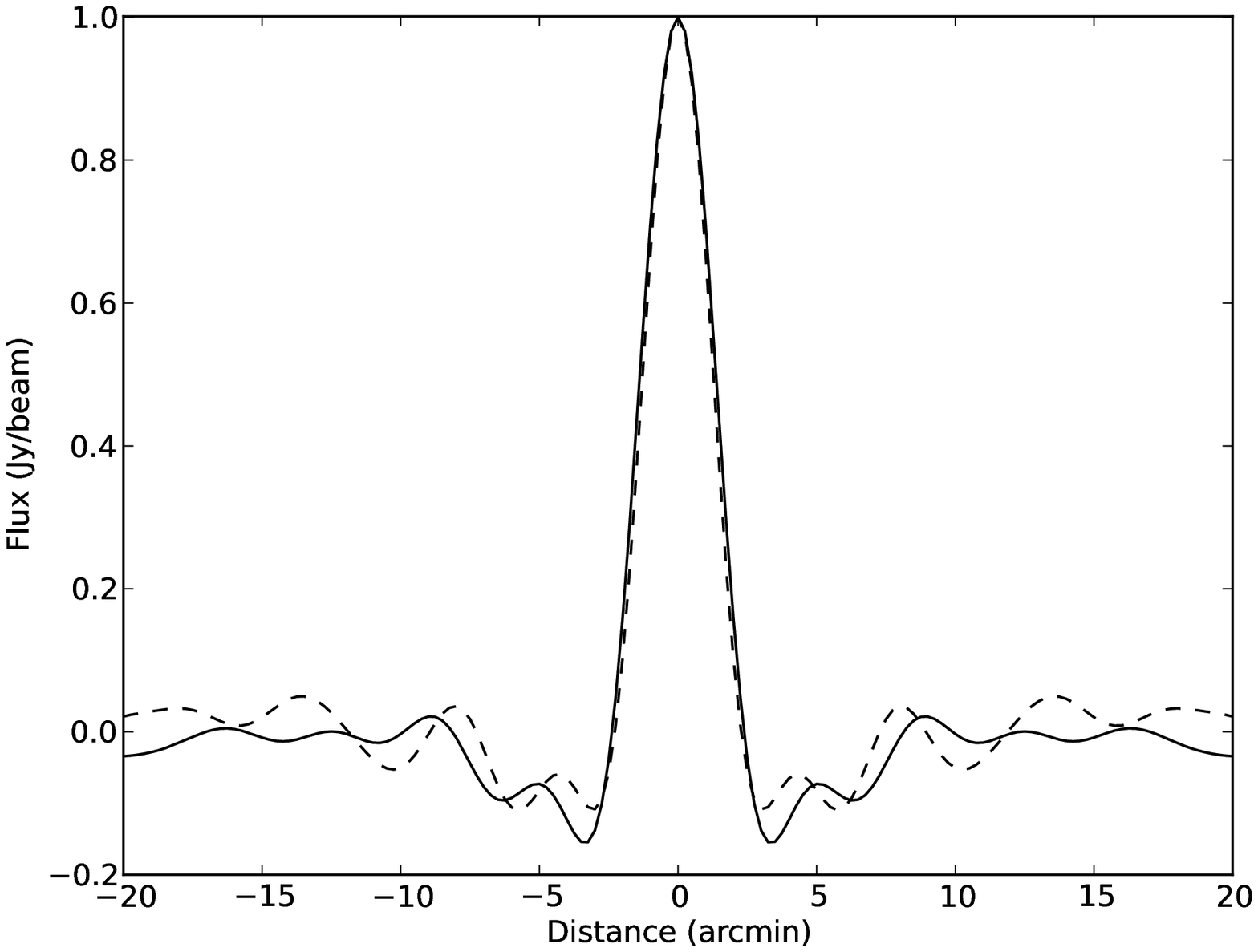}\qquad\includegraphics[width=5.0cm,height= 5.0cm,clip=,angle=0.]{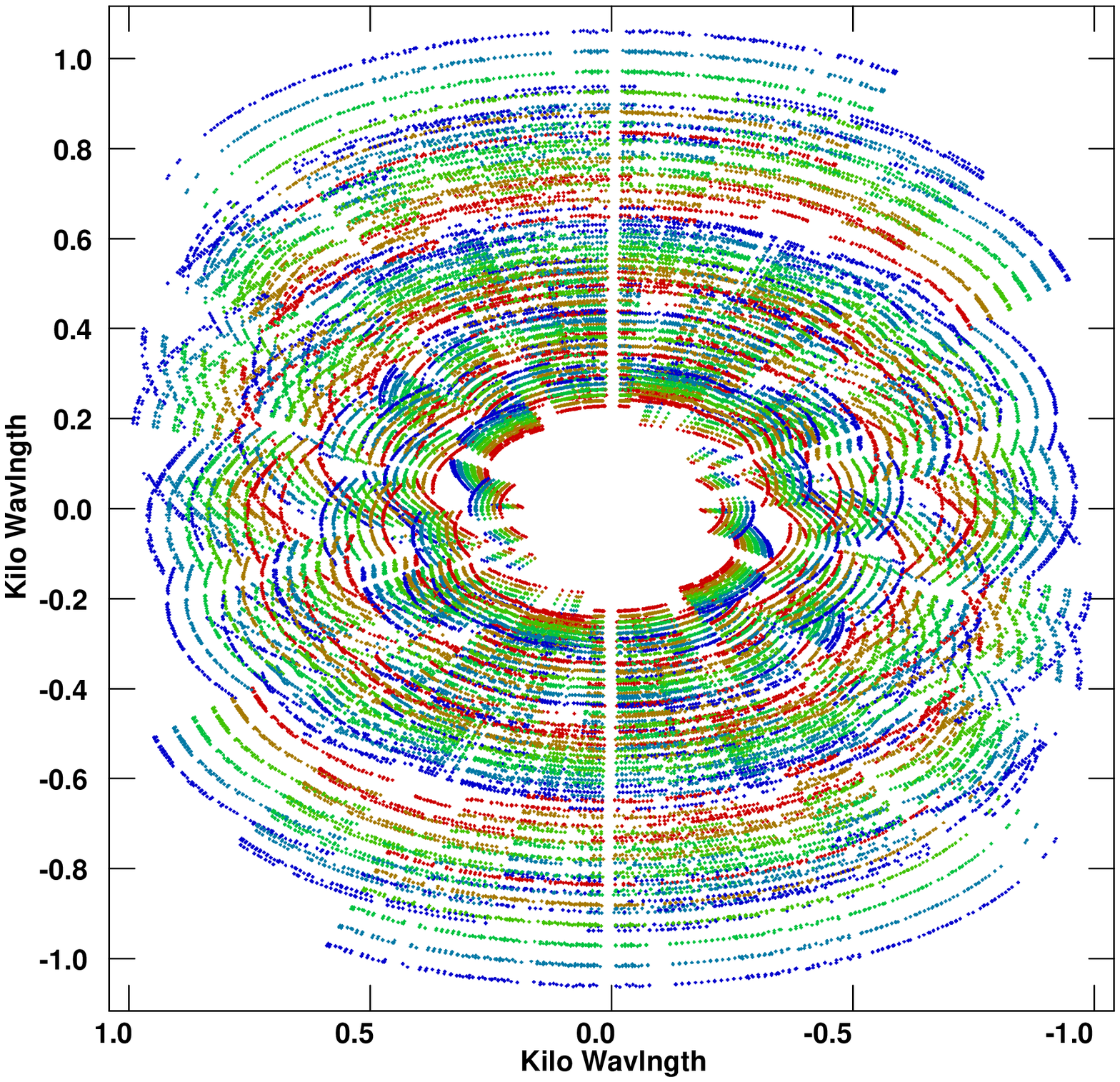}}
\caption{On the left is the synthesized beam for the AMI SA pointed observation towards the cluster (Figure \ref{fig:SA_pointed}). The contour levels range from 6\% to 30\% with intervals of 3\%; positive contours are solid lines and negative contours are dashed lines. The image in the centre shows one-dimensional slices through the centre of the synthesised beam; dashed lines show the profile along the declination axis and solid lines show the profile along the right ascension axis. The image on the right shows the corresponding  \textit{uv} coverage of the observation; a different colour is used for each AMI channel.
\label{fig:SA_pointed_syn}}
\end{figure*}

The cluster candidate was followed up with a pointed
observation. Within the 10\% point of the SA power primary beam 31 sources
were observed with the LA, 9 of which were detected at above 4
$\sigma_{\rm{SA,pointed}}$ and are modelled by our Bayesian
analysis. These 9 sources are a subset of the 16 sources modelled on
the Bayesian analysis of the survey data; they are indicated by a
`tick' in the last column of Table \ref{AMI002_sources}.  We find no
evidence that any of these 9 sources is extended relative to the LA
synthesised beam.  The image produced from the pointed-observation
data is shown before and after source subtraction in Figure
\ref{fig:SA_pointed}. Again we see a highly-extended, non-circular
negative feature with a peak flux-density decrement of $\approx$ 0.6
mJy ($8\sigma_{\rm{SA,pointed}}$). The SA synthesized beam and the
\textit{uv} coverage for the follow-up pointed observation is shown in
Figure \ref{fig:SA_pointed_syn}.  To estimate the maximum level of
contamination from the residuals of the sources in Table
\ref{AMI002_sources}, we assume that the residual is equal to the
error in the source flux and sum the absolute value of the synthesized
beam contribution from each of these residuals at the positions of
candidate 1 and candidate 2, we find contributions of 32$\mu \rm{Jy}$
and  70$\mu \rm{Jy}$ respectively. Hence, if in the unlikely case all
sources leave a feature of magnitude equal to the error in that source
flux, and that these features conspire in such a way to contribute
only negative flux at the positions of candidates 1 and 2, we find the contribution to the total SZ signal is minimal. This calculation does not account for any errors in the shape of the synthesized beam, due to e.g. antenna positions.

Our Bayesian analysis of the pointed-observation data, which have a
higher signal-to-noise ratio than the survey data, finds {\em two}
local peaks in the marginalised posterior distribution in the
$(x_c,y_c)$-plane. These cluster candidates are: candidate 1 at
J~03$^{\rm{h}}$~00$^{\rm{m}}$~14.8$^{\rm{s}}$
+26$^\circ$~10$'$~02.6$''$ and candidate 2 at
J~03$^{\rm{h}}$~00$^{\rm{m}}$~08.9$^{\rm{s}}$
+26$^\circ$~16$'$~29.1$''$. The Model (1) significance of the two cluster
detections are
$R_1 \approx 7.9 \times 10^{4}$ and $R_2 \approx 560$ respectively when we
apply the \cite{Evrard_02} model and $R_1 \approx 2.1 \times 10^{5}$
and $R_2 \approx 1800$
respectively when we apply the \cite{Jenkins} model.
These larger values for the $R$-ratio, as compared with those obtained
using the survey data, result from the higher signal-to-noise ratio
of the pointed observation.  The evidence values, $R$-ratios and
related parameters for the survey observations and the pointed
observation are summarised in Table \ref{tab:McAdam_evidences}. We
also made a direct comparison of the Bayesian evidence for a model
containing two clusters and a model containing just a single cluster
and find that the Bayesian evidence is 7.6 $\times$ $10^{5}$ higher
for the model containing two clusters.

\begin{table*}
\caption{Evidences, R-ratios and related parameters for the detection
  of the cluster candidates in the triangle of survey observations and
  the follow-up deep pointed observation. The limiting total mass is
  $M_{T,{\rm lim}} = 2\times 10 ^{14}$ $h_{70}^{-1}M_{\odot}$.}
 \label{tab:McAdam_evidences}
\begin{tabular}{lcccc}
 \hline
Parameter               & Survey   & Pointed (candidate 1)    & Pointed (candidate 2)   \\  \hline
Search area (steradians)& 2.00$\times$$10^{-5}$     &
2.35$\times$$10^{-5}$ &  2.35$\times$$10^{-5}$  \\ 
log($\tilde{Z}_{1,\rm{Jenkins}}$)   & 56351.1  & 29692.1    & 29687.3    \\ 
log($\tilde{Z}_{1,\rm{Evrard}}$)    & 56350.9  & 29692.0    & 29687.1    \\ 
log($\tilde{Z}_{0}$)           & 56346.6  & 29678.8   & 29678.8  \\ 

$\mu_{s,\rm{Evrard}}$        & 0.11     &  0.14    &  0.14     \\  
$\mu_{s,\rm{Jenkins}}$       & 0.29     &  0.34    &  0.34     \\ 
$R_{\rm{Press}}$             & 8.7      &  7.9$\times$$10^{4}$ &  560    \\ 
$R_{\rm{Jenkins}}$           & 26       &  2.1$\times$$10^{5}$  &  1800   \\ \hline
\end{tabular}
\end{table*}

\begin{figure*}
\centerline{\includegraphics[width=7.5cm,height= 7.5cm,clip=,angle=0.]{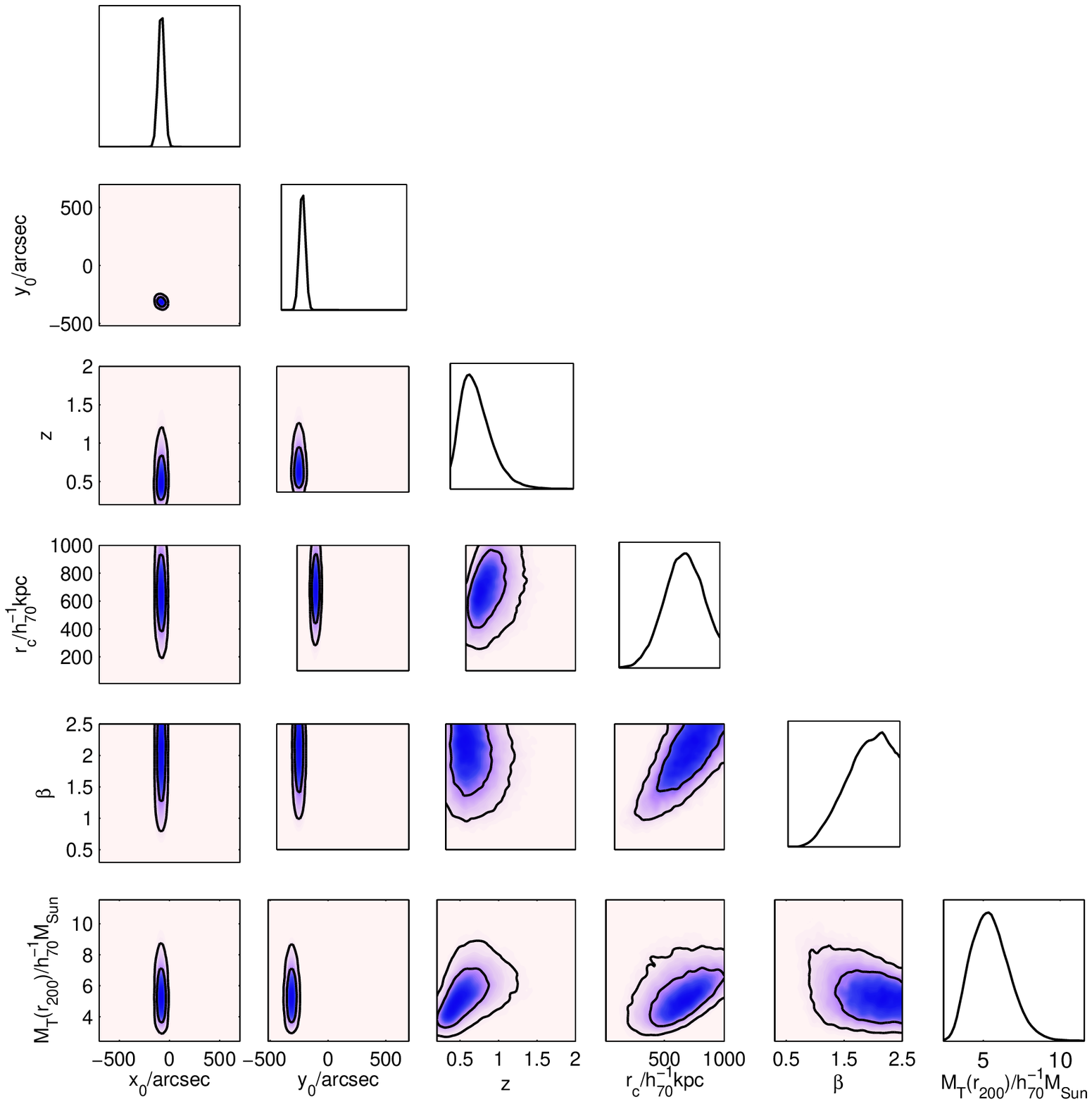}\qquad\includegraphics[width=7.5cm,height= 7.5cm,clip=,angle=0.]{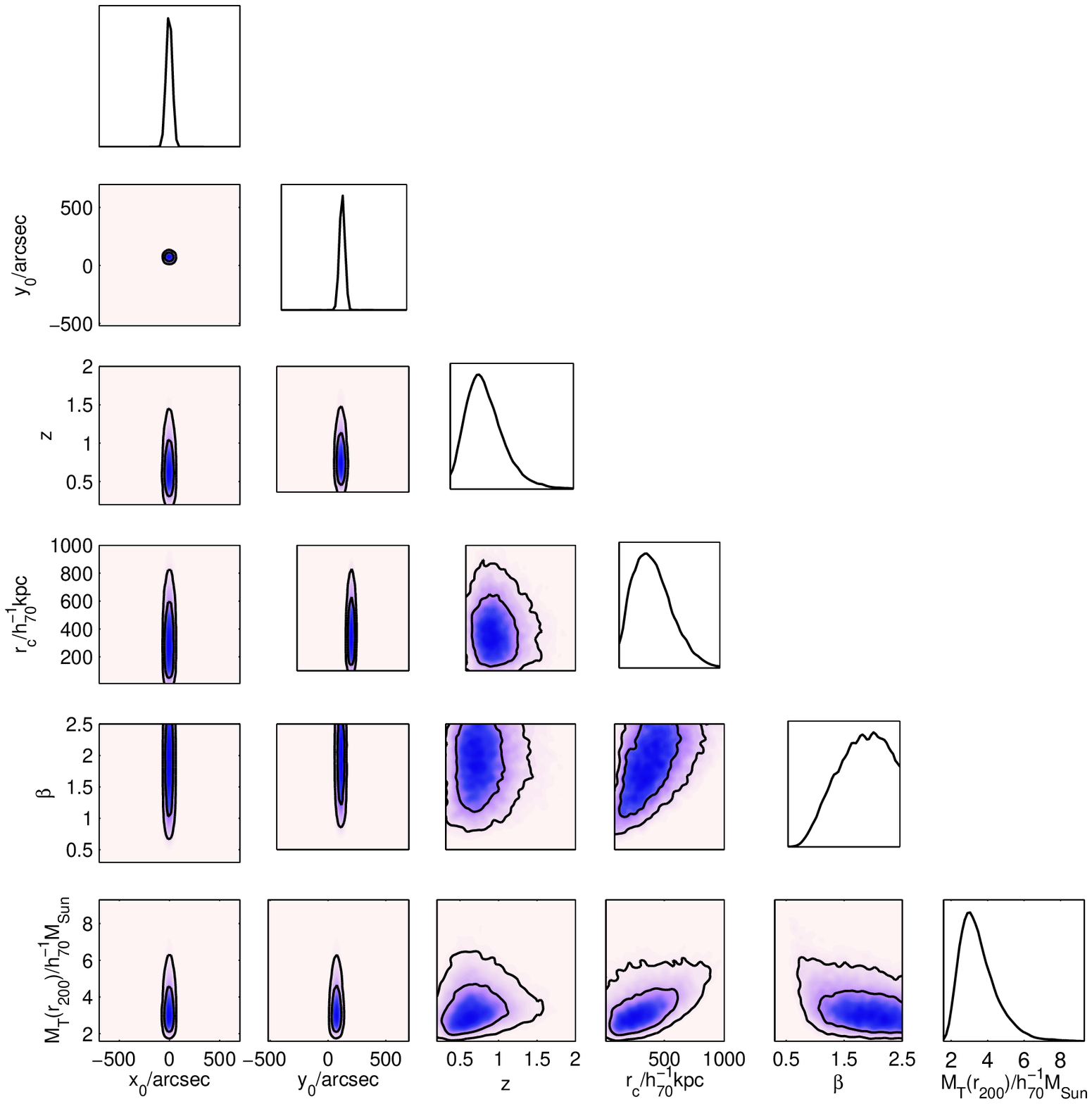}}
\caption{1D and 2D marginal posterior distributions for a selection of
  the parameters in physical cluster model for candidate 1 (left) and
  candidate 2 (right). The $M_{T,200}$ values have been divided by $10^{14}$.
\label{T323_tri}}
\end{figure*}

The 1D and 2D marginal posterior distributions for a selection of the
physical parameters of each cluster are shown in Figure~\ref{T323_tri}. We are able to constrain $M_{T,200}$, even though it is dengenerate with $z$. As $M_{T,200}$ is large this degeneracy causes the derived $z$ value to be low. We are able to constrain the well known degeneracy between $\beta$ and $r_c$ and find that values of $\beta$ $<$ 1.0 do not fit our data. We also find that the best-fit ratio of the lengths of the semi-minor to semi-major axes is 0.6 and 0.75 respectively; the orientation angles are 122$^\circ$ and 78$^\circ$.

\begin{figure*}
\centerline{\includegraphics[width=7.5cm,height= 7.5cm,clip=,angle=0.]{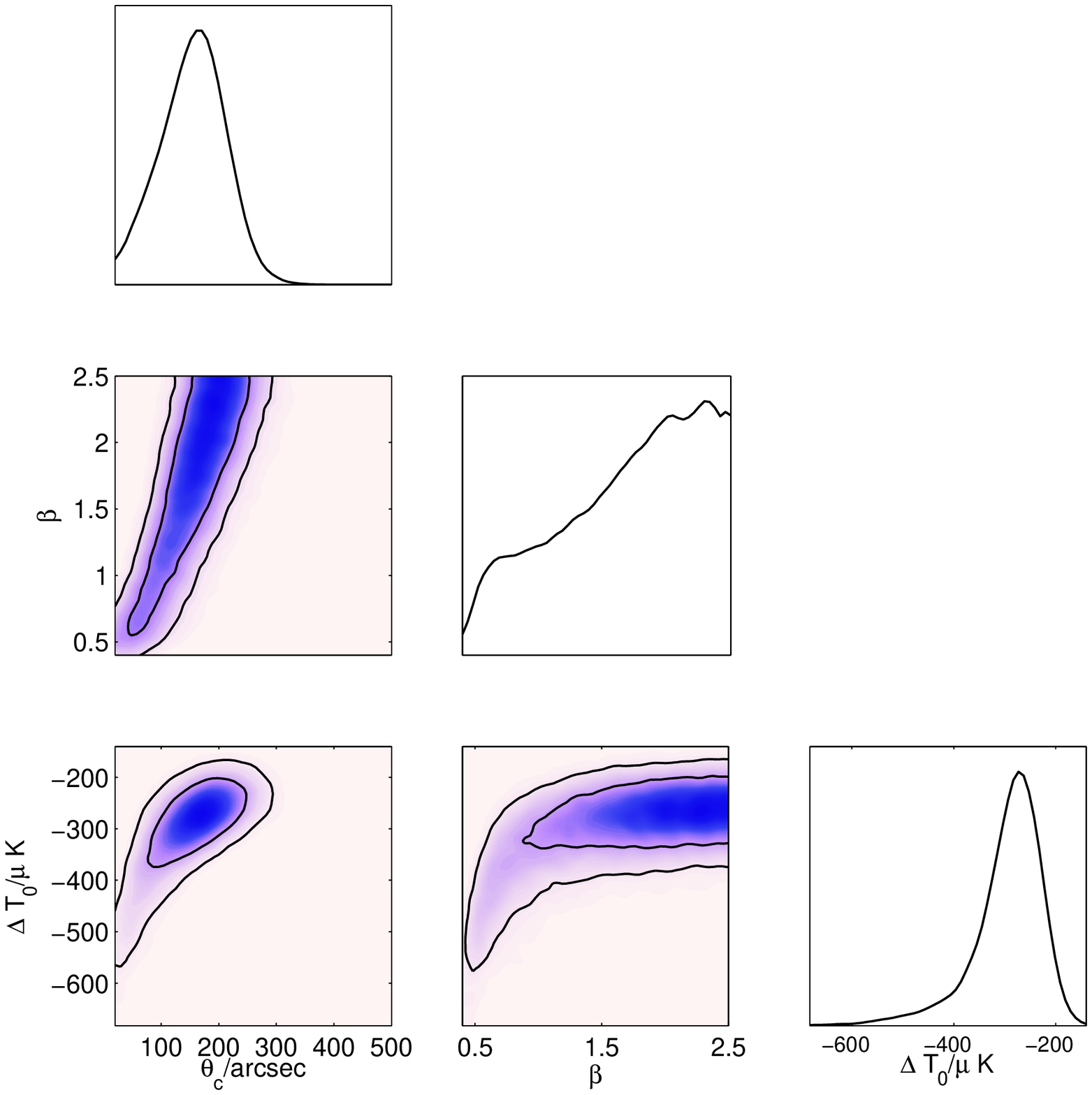}\qquad\includegraphics[width=7.5cm,height= 7.5cm,clip=,angle=0.]{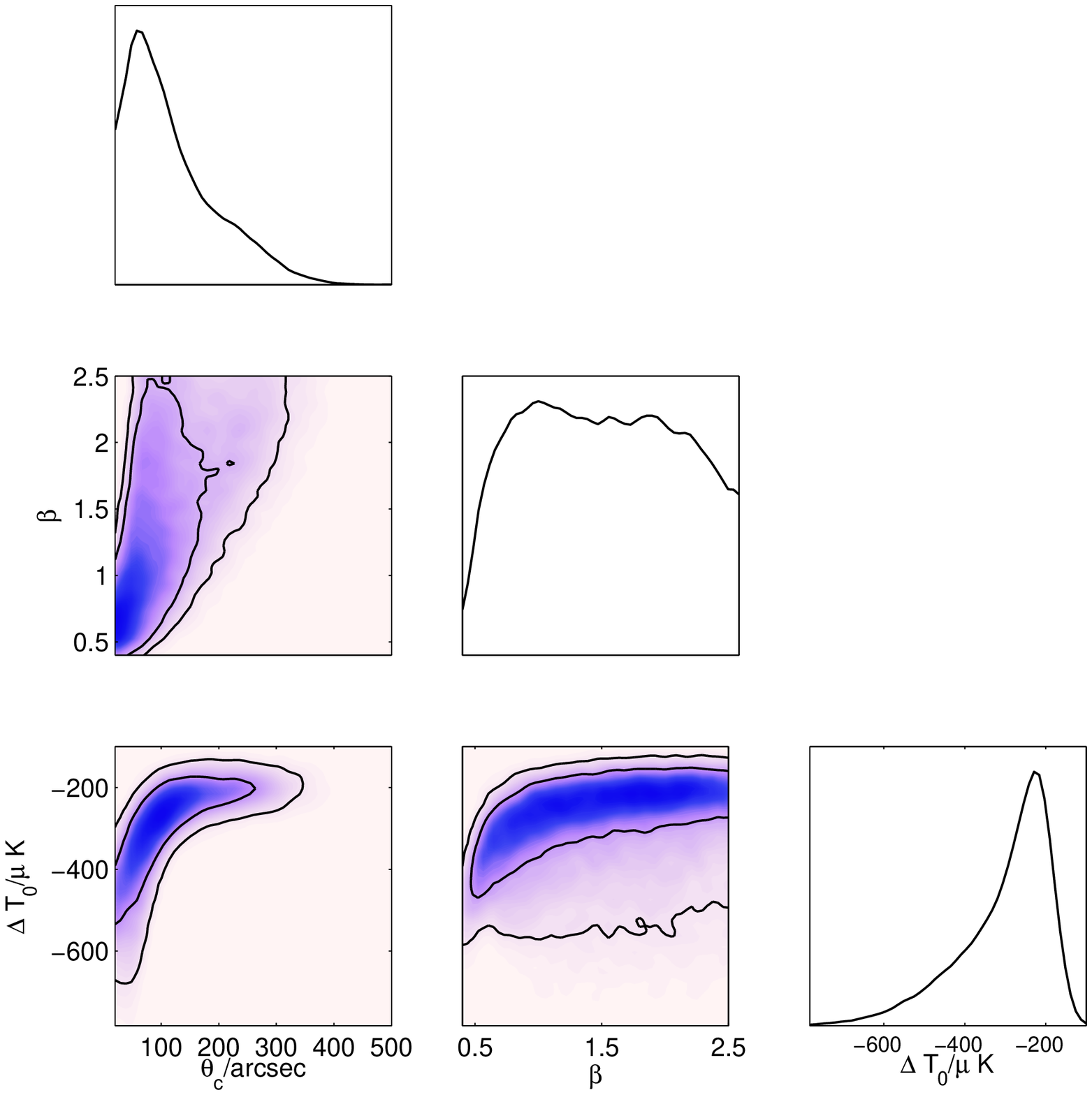}}
\caption{1D and 2D marginal posterior distributions for the
  parameters in the SZ decrement model for candidate 1 (left) and
  candidate 2 (right).} \label{fig:blob_plots}
\end{figure*}

Finally, we used Model (2) and performed a Bayesian analysis 
where we simply fit a $\beta$ profile to the SZ temperature decrement
directly. The 1D and 2D marginal posterior distributions for the
parameters $\theta_{c}$,
$\beta$ and $\Delta \rm{T_{0}}$ are shown in
Figure~\ref{fig:blob_plots}. From our data we are able to tightly constrain  $\Delta \rm{T_{0}}$ at $\approx -300\mu\rm{K}$, but we are unable to accurately derive $\beta$. The mean values and 68\% confidence limits
for each parameter are given in Table
\ref{tab:blob_results} and demonstrate significance of the detections
directly.

\begin{table}
\caption{Mean values and 68\% confidence limits for the parameters in
the SZ decrement model for candidate 1 and
  candidate 2.}
 \label{tab:blob_results}
\begin{tabular}{lcccc}
 \hline
Parameter               & Pointed (candidate 1)    & Pointed (candidate 2)   \\  \hline
$\theta_{c}/\arcsec$           & $156^{+27}_{-25}$    & $121^{+19}_{-100}$\\ 
$\beta$                       & $1.69^{+0.81}_{-0.24}$ & $1.46^{+1.03}_{-1.06}$\\ 
$\Delta T_{0}/\mu \rm{K}$          & $-295^{+36}_{-15}$   & $-302^{+70}_{-27}$\\ \hline
\end{tabular}
\end{table}

We have looked for optical identification of the cluster in the
Palomar all-sky survey and X-ray identification from
ROSAT \footnote{We have made use of the ROSAT Data Archive of the
  Max-Planck-Institut f$\rm{\ddot{u}}$r extraterrestrische Physik (MPE) at Garching,
  Germany.} -- no cluster identification is evident. We plan to perform X-ray and optical follow-up observations.

\section{Conclusions}\label{sec:CONC}

\begin{itemize}
\item We have presented a large, complex Sunyaev--Zel'dovich structure in an AMI blind field. The structure may be two separate components or be a single merging system.
\item A Bayesian analysis using a physical model for the cluster (including assumed priors on the number density of clusters) was used to constrain cluster parameters such as $\beta, r_{c}, M_{T,200}$ and $z$. Using the Bayesian evidences we have calculated formal probabilities of detection taking into account point sources, receiver noise and the statistical properties of the primary CMB anisotropy. For the deeper component we find a formal probability of detection ratio of  7.9 $\times$ $10^{4}$:1 when assuming the \cite{Evrard_02} cluster number count and 2.1 $\times$ $10^{5}$:1 when assuming \cite{Jenkins} as the true prior. We derive a cluster mass of ${M_{T,200}}=5.5^{+1.2}_{-1.3}$$\times$ $10^{14}{h_{70}^{-1}M_{\odot}}$. 
\item A Bayesian analysis using a phenomenological model of the gas distribution was also used to quantify the significance of the detection and again taking into account point sources, receiver noise and the statistical properties of the primary CMB anisotropy. For the deeper component we find $\Delta T_{0} = -295^{+36}_{-15} \mu \rm{K}$.
\item In our pointed follow-up observation  the cluster system is detected with a high significance, with each map indicating that there is a 0.6mJy/beam peak decrement ($8\sigma_{\rm{SA,pointed}}$) towards the deeper component and an integrated decrement flux density ($\rm{S_{SZ,integrated}}$) of $\rm{\approx1.2mJy\rm/beam}$. The other component has a
0.5mJy peak decrement and an integrated decrement of 0.7mJy.
\item Using the approximation $\rm{M_{T}^{5/3} \propto S_{\rm{SZ,integrated}}}$ we anticipate that the AMI blind cluster survey will detect clusters with $M_{T,200} >2$ $\times$ $10^{14}{h_{70}^{-1}M_{\odot}}$ at $4\sigma_{\rm{SA,survey}}$.

\end{itemize}

\section{ACKNOWLEDGMENTS}

We thank the anonymous referee for providing us with very constructive comments and suggestions. We thank PPARC/STFC for support of AMI and its operations. We are grateful to
the staff of the Cavendish Laboratory and the Mullard Radio Astronomy
Observatory for the maintenance and operation of AMI. CRG, MLD, MO, MPS, TMOF
and TWS acknowledge PPARC/STFC studentships. This work was performed using the
Darwin Supercomputer of the University of Cambridge High Performance Computing
Service (http://www.hpc.cam.ac.uk/), provided by Dell Inc. using Strategic
Research Infrastructure Funding from the Higher Education Funding Council for
England, and the Altix 3700 supercomputer at DAMTP, University of Cambridge
supported by HEFCE and STFC. We are grateful to Stuart Rankin and Andrey
Kaliazin for their computing assistance.

\bsp
\label{lastpage}


\begin{thebibliography}{}
\setlength{\labelwidth}{0pt}



\bibitem[\protect\citeauthoryear{AMI Consortium: Barker et al.}{2006}]{2006MNRAS.369.L1}
AMI Consortium: Barker R. W., et al., 2006, MNRAS, 369, L1

\bibitem[\protect\citeauthoryear{AMI Consortium: Davies et al.}{2009}]{DAVIES_09}
AMI Consortium: Davies M. L. D., et al., 2009, MNRAS, 400, 984

\bibitem[\protect\citeauthoryear{AMI Consortium: Davies et al.}{2011}]{DAVIES_10C} 
AMI Consortium: Davies M. L. D., et al., 2011, MNRAS, 415, 2708 

\bibitem[\protect\citeauthoryear{AMI Consortium: Franzen et al.}{2011}]{TMOF_10C} 
AMI Consortium: Franzen T.~M.~O., et al., 2011, MNRAS, 415, 2699 


\bibitem[\protect\citeauthoryear{AMI Consortium: Franzen et al.}{2009}]{TMOF_WMAP}
AMI Consortium: Franzen T.~M.~O., et al., 2009, MNRAS, 400, 995 

\bibitem[\protect\citeauthoryear{AMI Consortium: Hurley-Walker et al.}{2009}]{NHW_LA}
AMI Consortium: Hurley-Walker N., et al., 2009, MNRAS, 396, 365

\bibitem[\protect\citeauthoryear{AMI Consortium: Olamaie et al.}{2010}]{Olamaie_2010} 
AMI Consortium: Olamaie M., et al., 2010, arXiv, arXiv:1012.4996 

\bibitem[\protect\citeauthoryear{AMI Consortium: Rodr{\'i}guez-Gonz{\'a}lvez et al.}{2011}]{CARMEN} 
AMI Consortium: Rodr{\'i}guez-Gonz{\'a}lvez C., et al., 2011, MNRAS, 414, 3751 

\bibitem[\protect\citeauthoryear{Scaife et al.}{2008}]{AMI_SA_LDN1111}
AMI Consortium: Scaife A.~M.~M., et al., 2008, MNRAS, 385, 809

\bibitem[\protect\citeauthoryear{Scaife et al.}{2009}]{AMI_SA_LYNDS}
AMI Consortium: Scaife A.~M.~M., et al., 2009, MNRAS, 400, 1394 

\bibitem[\protect\citeauthoryear{AMI Consortium: Zwart et al.}{2008}]{2008MNRAS.391.1545Z}
AMI Consortium: Zwart J.~T.~L., et al., 2008, MNRAS, 391, 1545

\bibitem[\protect\citeauthoryear{AMI Consortium: Zwart et al.}{2010}]{7CLUSTERS}
AMI Consortium: Zwart J.~T.~L., et al., 2010, arXiv:1008.0443v2

\bibitem[\protect\citeauthoryear{Bartlett \& Silk}{1994}]{BART_SILK}
Bartlett J. G., Silk J., 1994, ApJ, 423, 12

\bibitem[\protect\citeauthoryear{Birkinshaw}{1999}]{BIRK_SZ_REVIEW}
Birkinshaw M., 1999, Phys. Rep., 310, 97

\bibitem[\protect\citeauthoryear{Birkinshaw, Gull \& Moffet}{1981}]{Birkinshaw_1981}
Birkinshaw M., Gull S. F., Moffet A. T., 1981, ApJ, 251, 69


\bibitem[\protect\citeauthoryear{Carlstrom, Holder \& Reese}{2002}]{CARL_SZ_REVIEW}
Carlstrom J. E., Holder G. P., \& Reese E. D. 2002, ARA\&A, 40, 643

\bibitem[\protect\citeauthoryear{Carlstrom, Joy \& Grego}{1996}]{CARL_JOY}
Carlstrom J. E., Joy M., Grego L., 1996, ApJ, 456, L75

\bibitem[\protect\citeauthoryear{Cavaliere \& Fusco-Femiano}{1976,1978}]{BETA_1}
Cavaliere A., Fusco-Femiano R., 1976, A\&A, 49, 137

\bibitem[\protect\citeauthoryear{Cavaliere \& Fusco-Femiano}{1978}]{BETA_2}
Cavaliere A., Fusco-Femiano R., 1978, A\&A, 70, 677

\bibitem[\protect\citeauthoryear{Evrard et al.}{2002}]{Evrard_02}
Evrard A. E., et al., 2002, ApJ, 573, 7

\bibitem[\protect\citeauthoryear{Feroz et al.}{2009}]{FF_MCADAM}
Feroz F., et al., 2009, MNRAS, 398, 2049

\bibitem[\protect\citeauthoryear{Feroz \& Hobson}{2008}]{MULTINEST_1}
Feroz F., Hobson M P., 2008, MNRAS 384, 499

\bibitem[\protect\citeauthoryear{Feroz, Hobson \& Bridges}{2008}]{MULTINEST_2}
Feroz F., Hobson M. P., Bridges M., 2008, MNRAS 398, 1601


\bibitem[\protect\citeauthoryear{Grainge et al.}{1996}]{RT_GRAINGE}
Grainge K., Jones M., Pooley G., Saunders R., Baker J., Haynes T., Edge A., 1996, MNRAS, 333, 318


\bibitem[\protect\citeauthoryear{Halverson et al.}{2009}]{APEX_HALV}
Halverson, N. W., et al. 2008, ApJ, 701, 42

\bibitem[\protect\citeauthoryear{High et al.}{2010}]{SPT_CLUSTERS3}
High F.~W., et al., 2010, ApJ, 723, 1736 


\bibitem[\protect\citeauthoryear{Hincks et al.}{2010}]{ACT_CLUSTERS}
Hincks A.~D., et al., 2010, ApJS, 191, 423 


\bibitem[\protect\citeauthoryear{Hobson \& Baldwin}{2004}]{Metromod}
Hobson M. P., Baldwin J. E., 2004, Applied Optics, 43, 2651.

\bibitem[\protect\citeauthoryear{Hobson \& Maisinger}{2002}]{MH_MCADAM}
Hobson M. P., Maisinger K., 2002, MNRAS, 334, 569

\bibitem[\protect\citeauthoryear{Holler et al.}{2007}]{Holler_2007} 
Holler C.~M., Kaneko T., Jones M.~E., Grainge K., Scott P., 2007, A\&A, 464, 795 

\bibitem[\protect\citeauthoryear{Holzapfel et al.}{1997}]{SuZIE_HOL}
Holzapfel W. L., et al., 1997, ApJ, 479, 17

 \bibitem[\protect\citeauthoryear{Gordon \& Trotta}{2007}]{Gordon_2007} 
 Gordon C., Trotta R., 2007, MNRAS, 382, 1859 

 \bibitem[\protect\citeauthoryear{Jeffreys}{1961}]{Jeffreys_1961} 
Jeffreys, H., 1961, Theory of probability, Clarendon Press, Oxford, 3rd ed., 50, 69

\bibitem[\protect\citeauthoryear{Jenkins et al.}{2001}]{Jenkins}
Jenkins A., et al. 2001, MNRAS, 321, 372

\bibitem[\protect\citeauthoryear{Jones et al.}{1993}]{Jones_1993}
Jones M. E., et al. 1993, Nature, 365, 320

\bibitem[\protect\citeauthoryear{Kneissl et al.}{2001}]{AMI_EXPECTED_RESULTS}
Kneissl R., et al. 2001, MNRAS, 328, 783

\bibitem[\protect\citeauthoryear{Komatsu et al.}{2010}]{WMAP_FGAS_VALUE}
Komatsu E., et al., 2010, arXiv:1001.4538 

\bibitem[\protect\citeauthoryear{Kosowsky}{2003}]{ACT_KOSO}
Kosowsky, A., et al. 2003, New Astronomy Reviews, 47, 939

\bibitem[\protect\citeauthoryear{Lahav et al.}{2002}]{2DF_SIGMA8}
Lahav O. et al. 2002, MNRAS, 333, 961


\bibitem[\protect\citeauthoryear{Lancaster et al.}{2005}]{VSA_LANC}
Lancaster K., et al. 2005, MNRAS, 359, 16


\bibitem[\protect\citeauthoryear{Lancaster et al.}{2007}]{OCRA_LANC}
Lancaster K., et al. 2007, MNRAS, 378, 673

\bibitem[\protect\citeauthoryear{Marshall et al.}{2003}]{MARSH_MCADAM}
Marshall P. J., Hobson M. P., Slosar A., 2003, MNRAS, 346, 489

\bibitem[\protect\citeauthoryear{Menanteau et al.}{2010}]{ACT_CLUSTERS2}
Menanteau F., et al., 2010, ApJ, 723, 1523 

\bibitem[\protect\citeauthoryear{Pearson et al.}{2009}]{CBI_PEAR}
Pearson T. J., et al, 2009, BAAS, 41, 447

\bibitem[\protect\citeauthoryear{Press \& Schechter}{1974}]{Press_Sch}
Press W. H., Schechter P., 1974, ApJ, 187, 425

\bibitem[\protect\citeauthoryear{Wu et al.}{2008}]{AMiBA_WU}
Wu J. -H. P., et al. 2009, ApJ, 694, 1619

\bibitem[\protect\citeauthoryear{Ruhl et al.}{2004}]{SPT_INTRO}
Ruhl J., et al. Proc. SPIE, Vol. 5498, p 11-19, 2004


\bibitem[\protect\citeauthoryear{Muchovej et al.}{2011}]{Muchovej_2011} 
Muchovej S., et al., 2011, ApJ, 732, 28 


\bibitem[\protect\citeauthoryear{Seljak et al.}{2005}]{SDSS_SIGMA8}
Seljak, U., et al., 2005, PhysRevD, 71, 103515

\bibitem[\protect\citeauthoryear{Staniszewski et al.}{2009}]{SPT_2009}
Staniszewski, Z., et al., 2009, ApJ, 701, 32

\bibitem[\protect\citeauthoryear{Sunyaev \& Zel'dovich}{1972}]{SZE}
Sunyaev R. A., Zel'dovich, Ya B., 1972, Comm. Astrophys. Sp. Phys., 4, 173

\bibitem[\protect\citeauthoryear{Udomprasert et al. }{2004}]{CBI_UDOM}
Udomprasert P. S., Mason B. S., Readhead A. C. S., Pearson T. J., 2004, ApJ, 615, 63

\bibitem[\protect\citeauthoryear{Vanderlinde et al.}{2010}]{SPT_CLUSTERS2}
Vanderlinde K., et al., 2010, ApJ, 722, 1180 

\bibitem[\protect\citeauthoryear{Vikhlinin et al.}{2009}]{CHANDRA_SIGMA8}
Vikhlinin A., et al., 2009, ApJ, 692, 1060

\bibitem[\protect\citeauthoryear{Voit}{2005}]{MT_scaling}
Voit G. M., 2005, Rev.Mod.Phys, 77, 207

\bibitem[\protect\citeauthoryear{Waldram et al.}{2010}]{LIZ_9C}
Waldram E. M., Pooley G. G., Davies M. L., Grainge K. J. B., Scott P. F, 2010, MNRAS, 404, 1005

\end{thebibliography}
\end{document}